\documentclass{aa}  

\usepackage{graphicx}
\usepackage{txfonts}
\usepackage{multirow}
\usepackage{pdflscape}
\usepackage{placeins}
\usepackage{float} 
\usepackage[breaklinks=true,hidelinks]{hyperref} 
\hypersetup{
  colorlinks=true,   
  urlcolor=cyan,     
  linkcolor=magenta,     
  citecolor=blue
}

\usepackage{longtable}

\newcommand{\gaia}{\textit{Gaia} }
\newcommand{\spitzer}{\textit{Spitzer} }
\newcommand{\hii}{H\textsc{ii} }
\newcommand{\logl}{$\textrm{log}(L/L_\odot)\,$}

\graphicspath{ {figs/} }

\begin{document} 

   \title{A machine-learning photometric classifier for massive stars in nearby galaxies}
   
   \titlerunning{A machine-learning classifier for massive stars in nearby galaxies II}

   \subtitle{II. The catalog}

   \author{G. Maravelias\inst{\ref{noa},\ref{forth}}
        \and
        A. Z. Bonanos\inst{\ref{noa}}
        \and
        K. Antoniadis\inst{\ref{noa}, \ref{nkua}}
        \and
        G. Mu\~{n}oz-Sanchez\inst{\ref{noa}, \ref{nkua}}
        \and
        E. Christodoulou\inst{\ref{noa}, \ref{nkua}}
        \and
        S. de Wit\inst{\ref{noa}}
        \and
        E. Zapartas\inst{\ref{forth}}
        \and
        K. Kovlakas\inst{\ref{ice}, \ref{ieec}}
        \and
        F. Tramper\inst{\ref{csic}}
        \and
        P. Bonfini\inst{\ref{alma}, \ref{uoc}}
        \and
        S. Avgousti\inst{\ref{nkua-dit}}
    }

   \institute{
        IAASARS, National Observatory of Athens, GR-15236, Penteli, Greece\label{noa}
        \and 
        Institute of Astrophysics, FORTH, GR-71110, Heraklion, Greece\label{forth}
        \and
        Department of Physics, National and Kapodistrian University of Athens, Panepistimiopolis, GR-15784, Zografos, Greece\label{nkua}
        \and
        Institute of Space Sciences (ICE), CSIC, Campus UAB, E-08193, Barcelona, Spain\label{ice}
        \and
        Institut d’Estudis Espacials de Catalunya (IEEC), Edifici RDIT, Campus UPC, E-08860, Castelldefels (Barcelona), Spain\label{ieec}
        \and
        Centro de Astrobiología (CSIC-INTA), E-28850, Torrejón de Ardoz, Spain\label{csic}
        \and
        Alma-Sistemi Srl, IT-00012 Guidonia, Italy\label{alma}
        \and
        Physics Department, and Institute of Theoretical and Computational Physics, University of Crete, GR-71003 Heraklion, Greece\label{uoc}
        \and 
        Department of Informatics and Telecommunications, National and Kapodistrian University of Athens, GR-16122, Greece\label{nkua-dit}
     }

   \authorrunning{Maravelias et al.}

  \date{Received XX XX, XXXX; accepted XX XX, XXXX}
 
  \abstract

\abstract
    {Mass loss is a key aspect of stellar evolution, particularly in evolved massive stars, yet episodic mass loss remains poorly understood. To investigate this, we need evolved massive stellar populations across various galactic environments.}
   {However, spectral classifications are challenging to obtain in large numbers, especially for distant galaxies. We addressed this by leveraging machine-learning techniques.}
   {We combined \spitzer photometry and Pan-STARRS1 optical data to classify point sources in 26 galaxies within 5 Mpc, and a metallicity range 0.07-1.36 Z$_\odot$. \gaia data release 3 (DR3) astrometry was used to remove foreground sources. Classifications are derived using a machine-learning model developed in our previous work.}
   {We report classifications for 1,147,650 sources, with 276,657 sources ($\sim24\%$) being robust. Among these are 120,479 red supergiants (RSGs; $\sim11\%$). The classifier performs well even at low metallicities ($\sim0.1$ Z$_\odot$) and distances under 1.5 Mpc, with a slight decrease in accuracy beyond $\sim3$ Mpc due to \spitzer’s resolution limits. We also identified 21 luminous RSGs (\logl$\ge5.5$), 159 dusty yellow hypergiants in M31 and M33, as well as 6 extreme RSGs (\logl$\ge6$) in M31, challenging observed luminosity limits. Class trends with metallicity align with expectations, although biases exist. }
   {This catalog serves as a valuable resource for individual-object studies and \textit{James Webb} Space Telescope target selection. It enables the follow-up on luminous RSGs and yellow hypergiants to refine our understanding of their evolutionary pathways. Additionally, we provide the largest spectroscopically confirmed catalog of extragalactic massive stars and candidates to date, beyond the Clouds, comprising 5,273 sources (including $\sim330$ other objects).}


   \maketitle

\section{Introduction}

One of the main goals of the \textit{James Webb} Space Telescope (JWST) is to study galaxies over cosmic time, from the early Universe to the present. However, their light is a combination of various components, one of which is their stellar content. Although rare in absolute numbers, massive stars make an important contribution. Through their feedback, whether from strong stellar winds or explosive supernovae, massive stars play a critical role in enriching and shaping the environments of their host galaxies. This is especially important in the early Universe, when metallicity was extremely low; the formation and evolution of such stars is still not well understood. The only way to gain insight into these objects is by examining these populations in nearby low metallicity galaxies. Resolved population studies in such galaxies are possible in the Local Group (e.g., Sextans A; \citealt{Lorenzo2022}, Large Magellanic Cloud (LMC); \citealt{xshootu_project}, Small Magellanic Cloud (SMC); \citealt{bloem_project}) but challenging at longer distances. Therefore, we lack well-explored populations of massive stars at these metallicities. 

The main goal of the ASSESS\footnote{\url{http://assess.astro.noa.gr/}} (Episodic Mass Loss in Evolved Massive stars: Key to Understanding the Explosive Early Universe) project \citep{Bonanos2024} was to investigate the role of episodic mass loss in the evolution of massive stars (see e.g., \citealt{Yang2023, Antoniadis2024, Antoniadis2025, deWit2024, Munoz-Sanchez2024, Munoz-Sanchez2025, Zapartas2025, Christodoulou2025}). A large number of sources with secure classifications was needed to explore its importance across various metallicity environments; therefore, we set up both an observing campaign to acquire spectra for a large number of sources (see e.g., \citealp{deWit2023, Bonanos2024, deWit2025}) as well as a machine-learning approach (\citealt{Maravelias2022}; henceforth Paper I). We developed a classifier that uses optical and IR photometry to select dusty, mass-losing, evolved massive stars and classify \spitzer-detected point sources into different broad classes. The purpose of this classifier was to predict the classes for approximately 1.2 M sources from 26 galaxies within 5 Mpc and spanning a metallicity range (0.07-1.36 Z$_\odot$), creating the largest point source catalog with spectral-type classifications for and beyond the Local Group. In Paper I, we provide the method and explore the classifier's prediction accuracy\footnote{Code and other material available at \url{https://github.com/gmaravel/pc4mas}.}, while the current paper presents the results of the classifier's application and the corresponding catalog of sources.

In Sect. \ref{s:data} we present the data collection and processing, the selection process of the best candidates (based on the results of the machine-learning classifier), as well as the collection of all sources with known spectral classification from the literature. In Sect. \ref{s:results}, we describe the catalog we compiled, along with statistics on the number of objects per class, and present exemplar color-magnitude diagrams (CMDs). In Sect. \ref{s:discussion}, we provide a comparison of the spectral types we predicted with those derived from the literature and discuss the performance of the classifier. We present the populations as functions of metallicity and explore the luminous red supergiants (RSGs) and dusty yellow supergiants (YSGs) found in our sample. Finally, in Sect. \ref{s:summary} we summarize and conclude our work.

\section{Data collection and processing}
\label{s:data}

In the following sections, we describe our sample and the steps we followed to build our catalogs and remove foreground sources. Although this has been extensively discussed in Paper I, we provide here a short overview, along with adjustments mainly in the foreground cleaning approach prompted by the release of \gaia data release 3 (DR3).

\subsection{Galaxy sample}

In Paper I, we used M31 and M33 to train the classifier, and WLM, Sextans A, and IC 1613 to test it. In this work, we present the results of applying the classifier to the whole sample of 26 galaxies included in the ASSESS project \citep{Bonanos2024} (except for the Clouds, which were treated separately; e.g., \citealt{Yang2019, Yang2020, Yang2021, Yang2023, deWit2023, Antoniadis2024, Munoz-Sanchez2024}).
This is presented in Table \ref{t:galaxy_properties}, along with the basic properties of the galaxies, including their names, coordinates, types, radii (corresponding to galaxy sizes, based on visual inspection, used to match the catalogs), distances, and metallicity.

\begin{table*}
  \centering
  \caption{Galaxies examined in this work along with some basic properties.}
  
   \label{t:galaxy_properties}
  \begin{tabular}{lcccrcr}
  \hline
  \hline
  Galaxy   & R.A. (J2000)    & Dec. (J2000)    & Type$^a$  & Radius & Distance$^b$ & Metallicity$^c$\\ 
       & (hh:mm:ss) & (dd:mm:ss) &       &  (\arcmin)     &  (Mpc)       & ($Z_{\sun}$) \\
  \hline
  WLM         & 00:01:58 & $-$15:27:39 & SB(s)m: sp     & 9   &  0.95$\pm$0.01 & 0.13$^1$\\
  NGC 55    & 00:14:54 & $-$39:11:48 & SB(s)m? edge-on            & 21      &  1.98$\pm$0.02             & 0.31$^2$ \\  
  IC 10     &   00:20:17 & +59:18:14 & dIrr IV/BCD    &  6  & 0.78$\pm$0.04  & 0.47$^3$ \\    
  M31     & 00:42:44 & +41:16:09 & SA(s)b LINER   & 105 & 0.75$\pm$0.02  & 1.36$^4$\\
  NGC 247  & 00:47:09 & $-$20:45:37 & SAB(s)d        & 14  & 3.56$\pm$0.03  & 0.54$^5$ \\
  NGC 253   & 00:47:33 & $-$25:17:18 & SAB(s)c        & 21  & 3.61$\pm$0.03  & 0.83$^6$\\
  NGC 300         & 00:54:53 & $-$37:41:04 & SA(s)d & 15      & 1.94$\pm$0.04  & 0.41$^7$ \\
  IC 1613   &   01:04:48 & +02:07:04 & IB(s)m         & 14  & 0.72$\pm$0.01  & 0.16$^8$\\
  M33         & 01:33:51 & +30:39:37 & SA(s)cd HII    & 30  & 0.85$\pm$0.02  & 0.65$^9$\\
  Phoenix Dwarf & 01:51:06 & $-$44:26:41 & IAm         & 8   & 0.43$\pm$0.01 & 0.07$^{10}$ \\
  NGC 1313  & 03:18:16 & $-$66:29:54 &  SB(s)d            & 8       & 4.21$\pm$0.06     & 0.35$^{11}$     \\
  NGC 2366  & 07:28:55 &  +69:12:57 & IB(s)m         & 6         & 3.21$\pm$0.04   & 0.16$^{12}$ \\  
  NGC 2403  & 07:36:51 & +65:36:09 & SAB(s)cd       & 14  & 3.13$\pm$0.06  & 0.56$^{13}$\\
  M81         & 09:55:33 & +69:03:55 & SA(s)ab        & 18  & 3.61$\pm$0.22  & $0.60^{14}$\\
  Sextans B   & 10:00:00 & +05:19:56 & IB(s)m         & 5   & 1.38$\pm$0.08    & 0.10$^{15}$ \\
  NGC 3109   & 10:03:07 & $-$26:09:35 & SB(s)m edge-on & 13  & 1.37$\pm$0.08  & 0.12$^{16}$\\
  NGC 3077    & 10:03:19  & +68:44:02  & I0 pec      & 5   & 3.75$\pm$0.11 & 0.89$^{17}$ \\
  Sextans A & 10:11:01 & $-$04:41:34 & IBm            & 4   & 1.41$\pm$0.05  & 0.07$^{15}$\\
  NGC 4214  & 12:15:39 & +36:19:37 & IAB(s)m        &  7  & 2.82$\pm$0.09  & 0.32$^{18}$\\
  NGC 4736  & 12:50:53 & +41:07:14 & (R)SA(r)ab     & 11   & 4.34$\pm$0.08  & 0.48$^{19}$\\
  NGC 4826 & 12:56:44   & +21:40:59  & (R)SA(rs)ab    & 7   & 4.36$\pm$0.06    & 0.22$^{20}$\\
  M83      & 13:37:01 & $-$29:51:56 & SAB(s)c        & 10  & 4.79$\pm$0.09  & 0.97$^{21}$\\
  NGC 5253  & 13:39:56  & $-$31:38:24 & Im pec        & 4   & 3.55$\pm$0.21     & 0.37$^{22}$\\
  NGC 6822  & 19:44:58 & +14:48:12 & IB(s)m         & 10  & 0.46$\pm$0.01  & 0.30$^{23}$\\
  Pegasus DIG & 23:28:36 & +14:44:35 & dI     & 8  &  0.95$\pm$0.03 & 0.10$^{24}$\\
  NGC 7793  & 23:57:50 & +32:35:28 & SA(s)d         & 8   & 3.39$\pm$0.06  & 0.48$^{25}$\\

  \hline
  \hline
  \end{tabular}
  \tablebib{
     $^1$\citet{Urbaneja2008}; 
     $^2$\citet{Hartoog2012}; 
     $^3$\citet{Cosens2024};
     $^4$\citet[][range 1.05-1.66 $Z_{\sun}$]{Zurita2012};
     $^5$\citealt{Davidge2021}, similar to LMC, 0.43 Z$_\sun$ from \citet{Hunter2007} and M33 - see below; 
     $^6$\citet{Spinoglio2022}; 
     $^7$\citet{Kudritzki2008}; 
     $^8$\citet{Bresolin2007};
     $^9$\citet[][range 0.3-1.0 $Z_{\sun}$]{U2009};
     $^{10}$\citet{Ross2015};
     $^{11}$\citet{Hadfield2007};    
     $^{12}$\citet{Thuan2005};    
     $^{13}$\citet[][range 0.35-0.77 $Z_{\sun}$]{Bresolin2022}; 
     $^{14}$\citet[][range 0.37-0.78 $Z_{\sun}$]{Arellano-Cordova2016}; 
     $^{15}$\citet[][range 0.07-0.14 $Z_{\sun}$]{Kniazev2005};
     $^{16}$\citet{Hosek2014}; 
     $^{17}$\citet{Storchi-Bergmann1994};
     $^{18}$\citet{Pilyugin2015};
     $^{19}$\citet{Moustakas2006};
     $^{20}$\citet[][from iron abundance, range 0.03-0.40 $Z_{\sun}$]{Kang2020};
     $^{21}$\citet[][range 0.35-1.58 $Z_{\sun}$]{Hernandez2019}; 
     $^{22}$\citet{Monreal2012};
     $^{23}$\citet{Patrick2015}; 
     $^{24}$\citet[][range 0.06-0.20 $Z_{\sun}$]{Skillman1997}
     $^{25}$\citet{DellaBruna2021}.
    }

   \tablefoot{   
   \tablefoottext{a}{Derived from NASA/IPAC Extragalactic Database (NED)}. \tablefoottext{b}{ From \cite{Tully2023}, except for M81, Sextans B, NGC 3109, and NGC 5253, which were derived from \cite{Tully2013}.}  \tablefoottext{c}{Metallicities are based on young clusters or massive stars or H~\textsc{ii} regions, when available. An average value was reported in cases where multiple measurements or radial gradients exist. For solar abundance we used 12+log(O/H) = 8.69 \citep{Asplund2009}.}
   
    }   

\end{table*}

\begin{figure}[H]
    \centering
    \includegraphics[width=\linewidth]{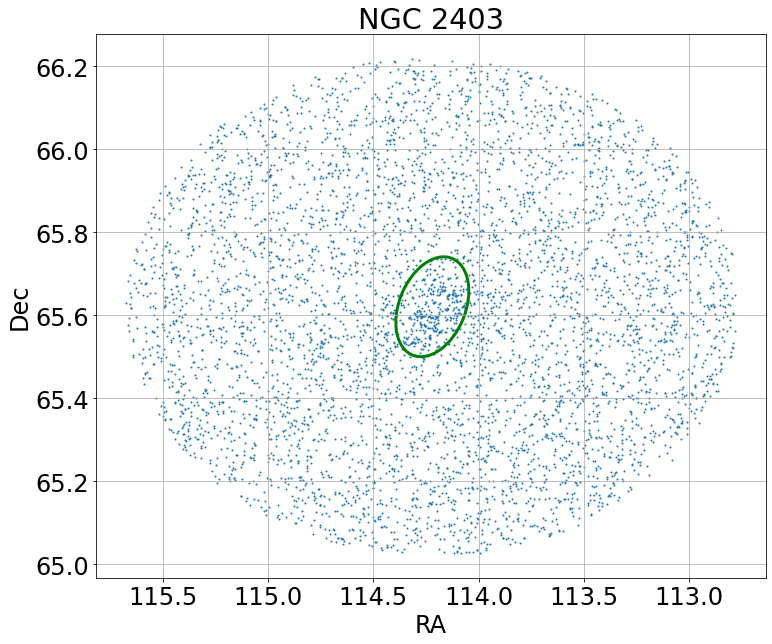}
    \includegraphics[width=\linewidth]{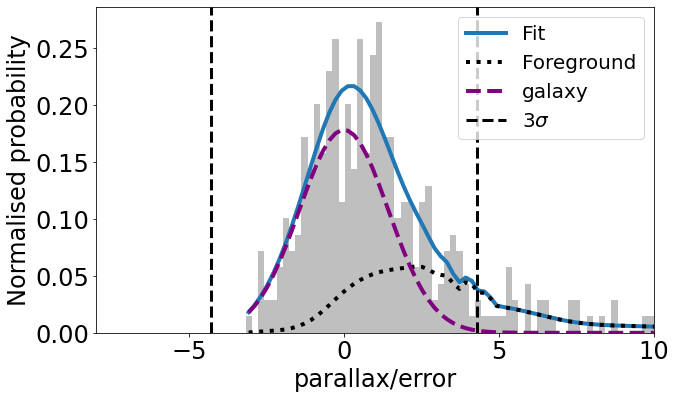}
    \includegraphics[width=\linewidth]{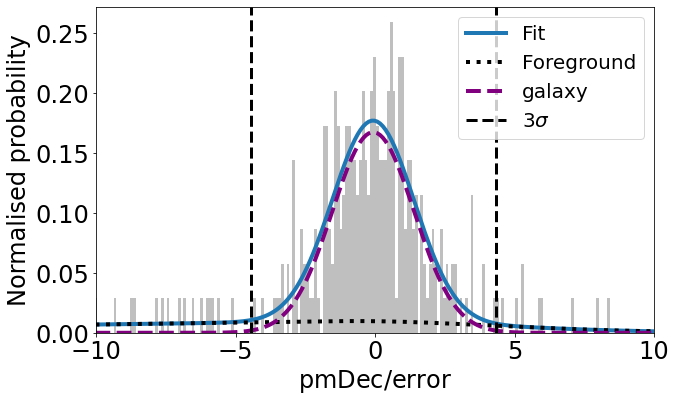}
    \caption{Example of fitting processing and foreground removal in NGC 2403. Top panel: \gaia sources where green ellipse defines  boundary selected for galaxy. Middle panel: Fitting parallax data for all sources within NGC 2403. The foreground contribution (dotted dark gray line) is scaled according to the densities of sources inside and outside the galaxy, while the dashed purple line shows the (Gaussian) contribution of galactic sources, and the blue line indicates the total fit. The vertical dashed black lines show the $3\sigma$ limits defined by the Gaussian properties of the galactic sources. Bottom panel: Similar to parallax, but for proper motion in declination (see Sect.~\ref{s:gaia} for more details). }
    \label{f:gaia_removal}
\end{figure}

\subsection{Surveys used and data processing}

Since IR photometry has been proven successful in identifying evolved massive stars \citep{Britavskiy2014,  Britavskiy2015, Britavskiy2019, Kourniotis2017} and those with dusty environments \citep{Bonanos2009, Bonanos2010, Bonanos2024}, we based our catalogs on precompiled point-source catalogs from the \spitzer Space Telescope (\citealt{Boyer2015, Khan2015, Khan2017, Williams2016}); (see Table \ref{t:galaxy_statistics} for details; the \spitzer column provides the number of sources within the radius as presented in Table \ref{t:galaxy_properties}). The mid-IR photometry ($\rm 3.6\, \mu m$, $4.5\, \mu \rm m$, $5.8\, \mu \rm m$, $8.0\, \mu \rm m$, $24\, \mu \rm m$\footnote{We are using the [3.6], [4.5], [5.8], [8.0], and [24] notation in the remainder of the paper.}) was crossmatched with optical photometry ($g,r,i,z,y$) obtained from the Panoramic Survey Telescope and Rapid Response System \citep[Pan-STARRS1;][]{Chambers2016}, or PS1, data archive, which is not available for the most southern galaxies. Additional photometry was obtained from the VISTA Hemisphere Survey (VHS; \citealt{McMahon2012}) and the UK Infra-Red Telescope (UKIRT) Hemisphere Survey (UHS; \citealt{Dye2018, Irwin2013}), however, with limited coverage of our sample for each survey\footnote{For only one case, NGC 5253, due to the lack of PS1 data and the availability of VHS, we used the \textit{y}-VHS instead.}.

\subsection{Removing foreground stars}
\label{s:gaia}

At the start of the project, \gaia data release 2 (DR2) was available. This release included a meaningful number of measurements (for parallax and proper motion) only for a few galaxies. Therefore, we initially performed foreground selection based on \gaia data for galaxies M31 and M33, to derive specific cuts in parallax and proper motion (see Paper I). With the release of DR3 \citep{Gaia2023}, both the amount of available data increased significantly and the quality of the measurements improved. Therefore, we decided to revisit this process. 

We performed foreground removal for each galaxy independently, defining parallax and proper motion limits based solely on sources within each galaxy. For this process, we defined an appropriate ellipse around each galaxy (by visual inspection of its optical image from the Digitized Sky Survey (DSS; red)) that enclosed the majority of its sources. This approach traces the spatial extent of resolved stellar populations providing a concrete and physically motivated boundary (in contrast to Petrosian and $D_{25}$ radii that are derived from integrated light profiles and brightness thresholds, respectively, and may not capture the true, sometimes fainter, outer edge\footnote{However, this approach is limited to galaxies with \gaia data and does not work for galaxies at larger distances for which the Petrosian and $D_{25}$ radii can provide homogeneous results.}). Additionally, we are interested in the statistical treatment of the data, so even though we may include larger radii in some cases, we are confident that we always properly sample the foreground population (which, in turn, is scaled for the size of the galaxy; see below). We next defined a search radius for \gaia sources around each galaxy by taking a radius about twice as large as the semi major axis of the ellipse defined around the galaxy (in contrast to the box defined around M31 and M33 in Paper I; however, the change does not affect the end result). We crossmatched (using a 1\arcsec\, search radius) the \gaia queried catalog around each galaxy with the corresponding catalog we had built from IR and optical photometry. Most galaxies have \gaia data, but this depends on the distance and crowding for each galaxy (the number of \gaia sources is presented in Table \ref{t:galaxy_statistics}).

Our purpose is to optimally select the foreground sources within the galaxy so that we eliminate (minimize, in reality) contamination, as foreground and galactic sources cannot be distinguished from the original IR catalogs. To achieve this, we followed the \gaia source selection from \citet{Lindegren2021}. For this, we first calculated the median error of each of the quantities \texttt{parallax\_error}, \texttt{pmra\_error}, and \texttt{pmdec\_error} and used this to replace all corresponding values that were smaller than their median values (as these might be underestimates of the real error). We excluded all sources without \gaia data and then selected sources with the following criteria: \texttt{astrometric\_excess\_noise}<1, \texttt{pmra\_error} / \texttt{pmdec\_error} <3,  \texttt{parallax\_error}<1.5, and  \texttt{phot\_g\_mean\_mag} $\le20.7$ (updated for DR3). Following Paper I, we first derived the distributions of the quantities \texttt{parallax/parallax\_error}\footnote{In Paper I we applied a 0.03 mas offset \citep{Lindegren2018}. However, it seems that its value is lower and possibly consistent with 0 \citep{Groenewegen2021}. In any case, the inclusion of this offset led to a subtle difference in the \texttt{parallax/parallax\_error} cut, which has no real impact in our work, and hence we did not apply it.}, \texttt{pmra/pmra\_error}, and \texttt{pmdec/pmdec\_error} for all sources outside the ellipse defined for each galaxy.
The distribution of the properties of these foreground sources was modeled using a regularized cubic spline (\texttt{scipy.interpolate.UnivariateSpline}, $k=3$) fitted to the histogram bin centers, with the smoothing parameter $s$ empirically adjusted to ensure smoothness without overfitting noise.
Then, we plotted the equivalent distributions for the sources within the ellipse, which consist of the foreground population and the true galactic one. The foreground contamination is estimated by the (previously defined) spline and scaled according to the ratio of the density inside and outside the galaxy, assuming that the foreground properties are approximately uniform across the field, as we estimated these properties from a region with a radius at least twice the size of each galaxy. For the galactic population, we assumed a Gaussian distribution. The total distribution was therefore fitted by combining this Gaussian along with the foreground contamination. After obtaining the mean value and standard deviation of this Gaussian profile, we determined as foreground sources those sources with values higher than $3\sigma$. For the \texttt{parallax/parallax\_error} we considered as foreground sources only those with values higher than $+3\sigma$ since negative values are not valid measurements. The selected values for each galaxy are presented in Table \ref{t:galaxy_statistics}. We also kept all sources without \gaia data, as there is no way we can decide whether such a source is foreground or galactic. 
By doing this, we actually increased contamination, but we were certain that we did not exclude possible interesting sources.

This step above was successful for the majority of the galaxies. However, there were cases where either the number of sources within the galaxy was too low (e.g., 17 stars in NGC 3077) or the distribution was very sparse and the fit was not trustworthy (e.g., IC~10 with 376 sources within the galaxy). For all these cases (11 out of 26 galaxies, indicated with an asterisk in their parallax and proper motion limits in Table \ref{t:galaxy_statistics}) we used the criteria derived from M31, which is the most populated galaxy. 

During crossmatching of the original IR catalogs with other surveys (including the \gaia DR2 initially), all multiple matches were removed (typically less than $\sim3\%$; Paper I). Upon inspection of the results, we found that a very small fraction of IR sources matched the same \gaia DR3 source. These sources were present in the initial point source catalogs with different IDs, but they had similar coordinates and magnitudes. We considered them as duplicates in the initial catalogs.
Although they matched the same \gaia source, the different magnitudes in IR could lead to different classification results. To avoid confusion, we opted to remove these sources altogether. Depending on the initial IR catalog, this fraction was less than $0.1\%$ (e.g., in NGC 4736 and M83 from \citealt{Khan2017}, or NGC~6822, NGC~300, M81 from \citealt{Khan2015}), to $\sim0.3\%$ (e.g., in NGC 253 and NGC 55; \citealt{Williams2016}) and up to $\sim7\%$ (e.g., in IC 10, IC 1613, and WLM; \citealt{Boyer2015}). 

Table \ref{t:galaxy_statistics} summarizes the initial number of \textit{Spitzer} sources; the crossmatches available with Pan-STARRS, \gaia\!, UHS, and VHS; parallax and proper motion limits for foreground detection (from \gaia\!); the number of sources selected based on these criteria, and the final number of sources after removing duplicates. This is the number of sources per galaxy analyzed in this work.

\subsection{Quality cuts}
\label{s:quality_cuts}

The application of the classifier was straightforward. After building a source catalog for each of our galaxies, we parsed it to the classifier, which identified and built the necessary features, i.e., calculated the color terms $r-i$, $i-z$, $z-y$, $y-[3.6]$, $[3.6]-[4.5]$, including missing data imputation if needed. This is possible for two reasons: first, because the compiled catalogs were drawn from independently curated datasets that already applied quality cuts; second, no preprocessing (e.g., signal-to-noise filtering) is necessary at this stage, since any low-confidence or noisy source can be removed at post-processing, as presented in this section. We used the pretrained models from Paper I to obtain the predictions for each galaxy\footnote{For the application of the classifier check the GitHub page of the project (\url{https://github.com/gmaravel/pc4mas}), where a python notebook and the models can be found.}. 

The classifier provides a classification among the following classes (see Paper I for details): BSG -blue supergiant, a rather loose class of early-type and hot stars; YSG - evolved yellow type stars; RSG - red supergiants; BeBR - B[e] supergiants; LBV - luminous blue variables; WR - Wolf-Rayet stars; GAL - point-like extragalactic sources (Active Galactic Nuclei (AGNs), Quasi-Stellar Objects (QSOs), compact galaxies), which enables the classifier to identify and remove background contaminants projected along the line of sight. The probability per class is provided for each of the algorithms used (i.e., support vector machines, random forest, multi-layer perceptron; see Paper I for more details), which are then  combined (with equal weighting) to provide the final set of probabilities. For each source, the final (predicted) class corresponds to the one with the highest probability of the ensemble approach (see Paper I for details).

In Paper I, we investigated the number of correct versus incorrect sources with probability. We found that the mean values for the correct and incorrect classifications (in the combined sample of M31 and M33) were $0.86 \pm 0.01$ and $0.60 \pm 0.03$, respectively. A cut at 0.86 significantly limits the sources to be considered, since the classifier performs slightly worse for lower metallicities than M31 and M33. Therefore, we opted to use a cut at 0.66, equivalent to the mean value $+3\sigma$ of the incorrect classification distribution, which gave us the opportunity to consider a larger sample of sources while keeping the fraction of incorrectly classified sources relatively low (cf., Figs. 6 and 8 of Paper I).

Additionally, band availability (i.e., missing data) significantly affects the performance and robustness of the prediction model, specifically probability. Our tests on missing data imputation (cf., Fig. 9 in Paper I) indicated that the classifier can effectively manage up to two missing features (equivalent to three missing bands, since each feature is a color index) with a loss of accuracy of approximately 10\%. This resilience extends to three missing features, albeit with an increased accuracy loss of less than 20\%. More missing values lead to an approximate 40\% loss, making the classifier unreliable. The reported accuracy losses correspond to the overall classifier performance, although individual classes are affected differently (cf., Fig. 12 in Paper I). However, class-specific adjustments are not feasible during the application of the classifier since the true class is unknown.

Given that missing values are a common occurrence for our sources, we identified two scenarios to select objects for further processing. The more balanced approach involved selecting sources with a final probability greater than 0.66 and band completeness exceeding 0.6 (i.e., missing two features, three bands in total). Alternatively, a more relaxed option considered sources with a probability greater than 0.50 and band completeness exceeding 0.4 (i.e., missing three features and four bands). However, the second scenario introduced a higher rate of false positives, increasing noise, due to more misclassifications, in the final results. Therefore, we decided to proceed exclusively with the first approach. To ensure flexibility, our published catalogs include data for all available sources, allowing users to apply their own selection criteria as needed.

\subsection{Collecting known sources from the literature}
\label{s:literature}

To better understand the strengths and limitations of our classifier, we needed to compare its predictions with previously classified sources. For this reason, we undertook the very demanding task of searching the literature for all known sources found in our galaxy sample. This task complements our work presented in Paper I,   where, during the development of the classifier, we were faced with many constraints imposed by \gaia DR2 coverage and smaller galaxy samples. We limited our search to sources with secured spectral classification, i.e., that had been obtained with spectroscopy and not from photometric or other (e.g., variability) criteria. We collected complete spectral samples - to the best of our knowledge - for all the 26 target galaxies, accounting for 5273 sources (from 83 different works). This number includes all known massive stars, including candidates, as well as another $\sim330$ point sources (such as carbon stars, background galaxies, \hii regions, planetary nebulae, and clusters). The numbers and corresponding references for each galaxy can be found in Table \ref{t:class_refs}. We note here that no spectral classifications were available for five galaxies (i.e., NGC 2366, NGC 3077, NGC 4214, NGC 4826, and NGC 5253). In all cases, we carefully checked for and removed duplicates, keeping the most recent and precise classifications. The current catalog serves as the most complete source of reference for spectral classifications for massive stars and candidates beyond the Milky Way and the Clouds.

\section{Results}
\label{s:results}

In this section, we provide the results of the application of the machine-learning classifier to our galaxy sample, in the form of a full catalog. This table contains the statistics per class (per galaxy) based on the most secure predictions. We also present and discuss some indicative CMDs derived directly from these results.

\subsection{Catalog description}

In Table \ref{t:thecatalog} we provide the first few lines of the final catalog for all sources from all galaxies, comprising 1,147,650 sources and spanning 78 columns. In the Table we provide the source ID number, \spitzer coordinates, \gaia DR3 ID number, proper motion and parallax, \spitzer\!\!, PS1, and near-IR photometry and errors,  previous classification (if available), probabilities per class for each method including the ensemble one, as well as final class, final probability, and band completeness.

\subsection{Populations}

\begin{table*}
  \centering
  \caption{Number of predictions per galaxy, after applying selection criteria on probability and band completeness (from Sect. \ref{s:quality_cuts}).} \label{t:galaxy_predictions}
  \begin{tabular}{l|r|rrrrrrr}
  \hline
  \hline
  Galaxy   &  Total  &  RSG   &  YSG  &  BSG  &  BeBR  &  WR   &  LBV  &  GAL  \\ 
  \hline

WLM & 526 & 268 & 85 & 13 & 2 & 147 & 1 & 10\\
IC 10 & 1622 & 658 & 11 & 0 & 0 & 947 & 0 & 6\\
M31 & 225176 & 81734 & 696 & 268 & 33 & 140153 & 12 & 2280\\
NGC 247 & 897 & 372 & 17 & 6 & 0 & 423 & 1 & 78\\
NGC 253 & 385 & 167 & 36 & 1 & 1 & 118 & 5 & 57\\
IC 1613 & 2964 & 2351 & 392 & 39 & 1 & 162 & 5 & 14\\
M33 & 31635 & 25808 & 322 & 212 & 23 & 4767 & 8 & 495\\
NGC 2366 & 42 & 29 & 5 & 0 & 0 & 2 & 0 & 6\\
NGC 2403 & 950 & 620 & 74 & 10 & 1 & 217 & 6 & 22\\
M81 & 1387 & 382 & 36 & 1 & 2 & 899 & 1 & 66\\
Sextans B & 231 & 176 & 25 & 2 & 0 & 24 & 0 & 4\\
NGC 3109 & 736 & 363 & 38 & 7 & 0 & 319 & 1 & 8\\
NGC 3077 & 96 & 17 & 1 & 0 & 0 & 68 & 0 & 10\\
Sextans A & 168 & 100 & 20 & 8 & 0 & 35 & 2 & 3\\
NGC 4214 & 142 & 42 & 10 & 4 & 0 & 81 & 0 & 5\\
NGC 4736 & 341 & 49 & 30 & 0 & 1 & 227 & 0 & 34\\
NGC 4826 & 151 & 24 & 19 & 0 & 0 & 94 & 0 & 14\\
M83 & 565 & 104 & 28 & 10 & 4 & 388 & 1 & 30\\
NGC 6822 & 8007 & 6992 & 220 & 34 & 4 & 690 & 17 & 50\\
Pegasus DIG & 627 & 217 & 15 & 1 & 0 & 389 & 0 & 5\\
NGC 7793 & 9 & 6 & 2 & 0 & 0 & 1 & 0 & 0\\
\hline
TOTAL  & 276657 & 120479 & 2082 & 616 & 72 & 150151 & 60 & 3197\\

  \hline
  \hline
  \end{tabular}

\end{table*}

For further consideration and exploration of the results, we only kept sources that satisfied the quality criteria as defined in Sect. \ref{s:quality_cuts}. This excluded the most uncertain predictions from the classifier which add noise, either due to the low probability or the number of missing values (below the band completeness threshold we set). Consequently, it allowed us to investigate the results, and assess the performance of the classifier. Table \ref{t:galaxy_predictions} presents the total number of source classifications, along with the number of predictions for the classes of RSG, YSG, BSG, BeBR, WR, LBV, and GAL, independently. 

Of the 26 galaxies, five (namely NGC 55, NGC 300, Phoenix Dwarf, NGC 1313, and NGC 5253) have zero sources (hence are not presented in Table \ref{t:galaxy_predictions}). This is due to a lack of PS1 data, which leads to more missing values than we accepted (less than 0.6). Since only the IR data are available for these galaxies, the predictions were highly uncertain and were therefore excluded from further analysis. 

Overall, we find that the highest number of predictions is for RSGs and WR stars. In particular, M31, M33, NGC 6822, IC 1613, IC10, and M81 have more than 1000 classified sources. However, we already know (from Paper I) that WR predictions suffer from many false positives. We further discuss the performance of the classifier in Sect. \ref{s:discuss_literature}. 

\subsection{Color-magnitude diagrams}

\begin{figure*}[htbp]
    \centering
    \includegraphics[totalheight=11.3cm]{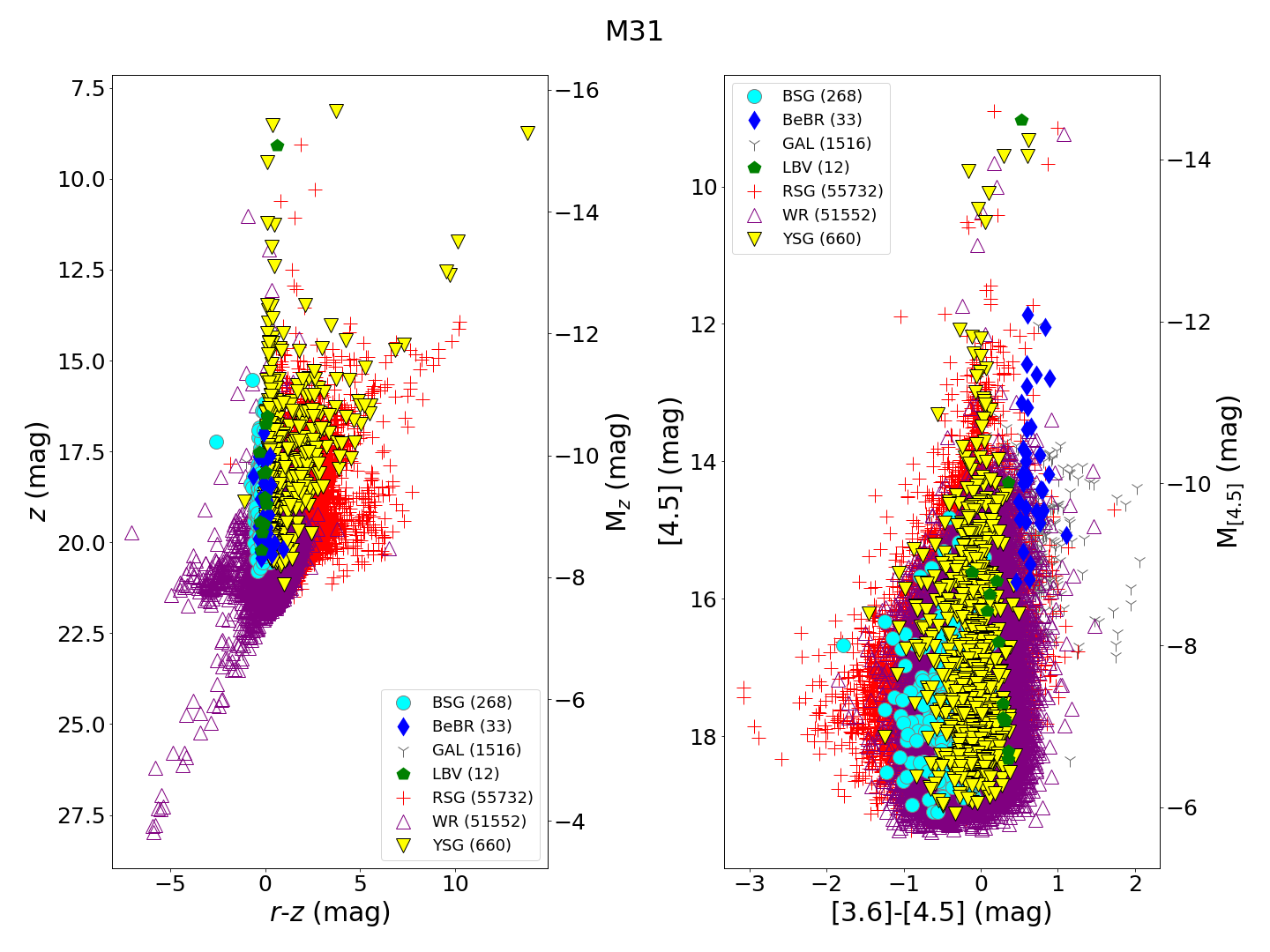}   \includegraphics[totalheight=11.3cm]{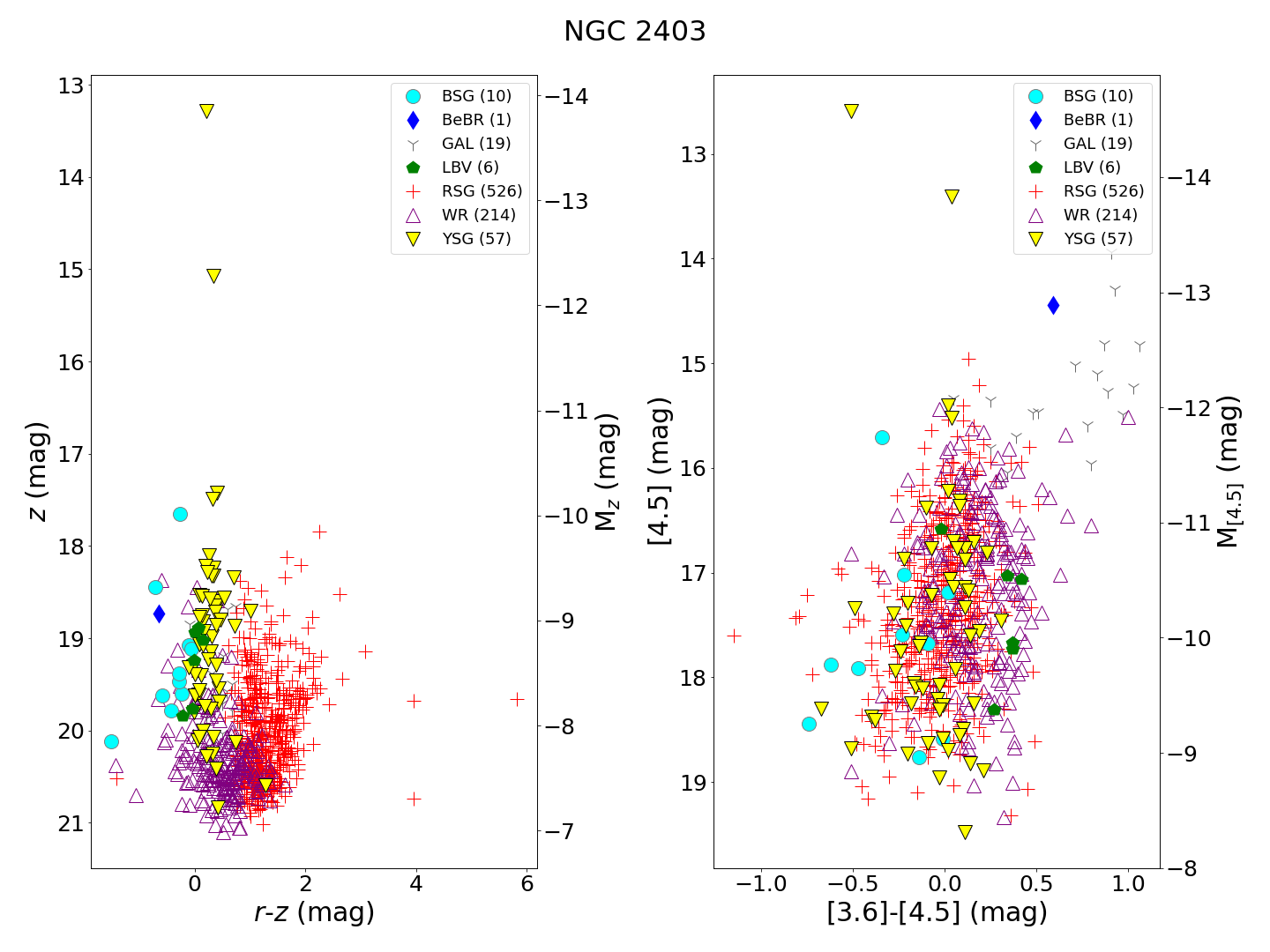}
    \caption{Example of CMDs ($z$ vs. $r-z$ and [4.5] vs. $[3.6]-[4.5]$) for M31 (top) and NGC~2403 (bottom) with predicted classifications. We plot all sources that satisfy the quality cuts imposed in Sect. \ref{s:quality_cuts}, and their photometric values come from the original data (i.e., we do not plot sources whose values have been imputed during the application of the classifier). We use cyan circles for blue supergiants (BSG), blue diamonds for B[e] supergiants (BeBR), gray triangles for galaxies (GAL), green pentagons for luminous blue variables (LBV), red crosses for red supergiants (RSG), purple empty triangles for Wolf-Rayet stars (WR), and yellow filled bottom-sided for yellow supergiants (YSG). The total number of sources per class is provided next to the class in the legend.}
    \label{f:cmds}
\end{figure*}

We explored the locations of the predicted populations in CMDs. Figure \ref{f:cmds} shows optical and mid-IR CMDs for two example galaxies, M31 (as the most populated galaxy) and NGC~2403.
We note that we do not plot sources whose values have been imputed during the application of the classifier, but only sources with their original photometry. We use different symbols to indicate the predicted class for each source and show the total number of predictions per class in the legend. 

We find that the positions of the sources match their predictions. In the $z$ versus $r-z$ CMD we see BSG located on the left (of approximately $r-z\sim0$) while YSG and RSG extend to redder colors as expected. The few predicted BeBR and LBV  are (consistently) located close to the BSG and WR. The latter display a much broader distribution, which is not real in many cases, as this class suffers the most from false positives. This is more striking in the case of M31, where we note a "tail" of WR that extends from the bulk of sources (with $r-z\sim0$ mag) to a point of $r-z\sim-5$ mag (and of $z\sim27.5$ mag). We attribute this trend to a combination of faint-end detection bias and color-space overlap between the WR and other classes (including additional faint blue populations visible at the distance of M31, for which the classifier is not trained), rather than from systematically noisy photometry. In contrast, we do not see this in NGC~2403. 

Regarding the [4.5] versus $[3.6]-[4.5]$ CMD we notice that the majority of RSGs lie around $[3.6]-[4.5]\sim0$ mag in both M31 and NGC~2403, consistent with what we would expect. For M31 though, there are sources with bluer colors. Although there are indeed RSGs with bluer colors (e.g., \citealt{Bonanos2009, deWit2024}), there is confusion with other populations (due to the proximity of M31). There is definitely confusion of WRs with other populations (WR is the class with the highest false positive results). In NGC~2403  WRs extend to slightly redder colors, which is consistent with what we expect. Yellow supergiants are located around 0.0 for both galaxies. However, the two very bright points in NGC~2403 (sources NGC2433-37 and NGC2433-66) are probably foreground sources, as there are no \gaia data that can confirm their nature. In this CMD, GAL seems to become more evident from the bulk of the stellar populations, and they are distinctly separated with $[3.6]-[4.5]>0.5$ mag. 

Overall, the position of the sources matches their predicted classification. Of course, there are a fraction of sources that are misclassified. This is due to a number of physical and technical reasons, such as the fact that strict boundaries in these parameter spaces between some classes do not exist (e.g., BSG and WR, YSG, and RSG), the fact that there could still be some contaminant sources, either because these have not been part of the classifier's training or because there could be photometric errors (e.g., crowding, aperture contamination, or remaining foreground sources). We further explore these in Sect. \ref{s:discuss_literature}.

\section{Discussion}
\label{s:discussion}

In this section, we compare our predictions with the literature data to understand the performance of the classifier. We also explore the trends of the various classes with metallicity, and discuss the luminous RSGs and dusty YSGs in our sample.

\subsection{Performance: comparison to the literature}
\label{s:discuss_literature}

\begin{figure*}[htbp]
    \centering
    \includegraphics[width=\textwidth] {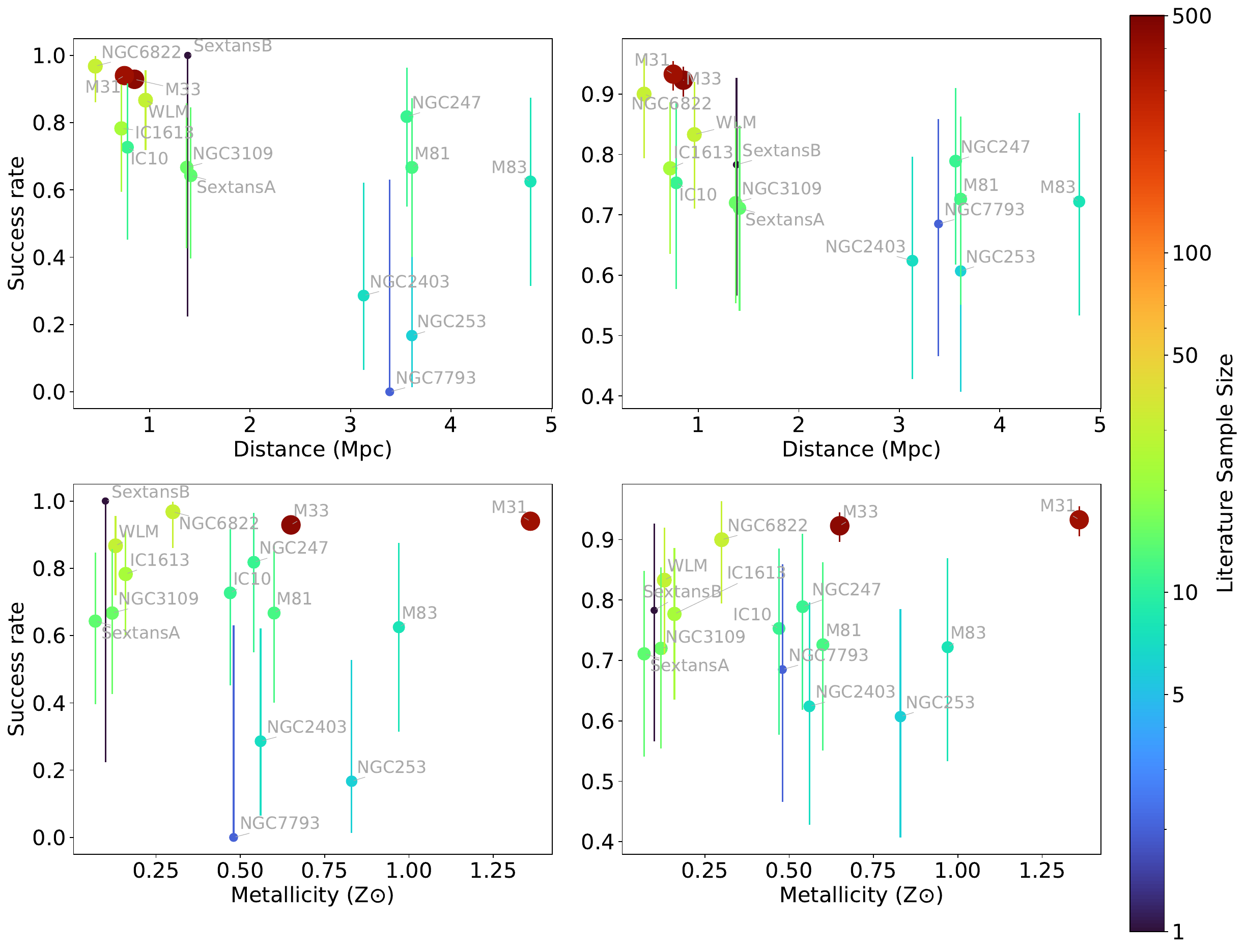}
    \caption{Success rate vs. distance (top panels) and metallicity (bottom panels), using a uniform prior (left panels) and a unimodal beta distribution (right panels) with a peak corresponding to $77\pm7$\% (based on the performance of the classifier during development). We notice a small decrease in the success rate with distance and a relatively flat behavior with metallicity, especially in the case where a prior is implemented (see Sect. \ref{s:discuss_literature} for details). The number of available classified sources from the literature is indicated by the size of the points and the colorbar on the right. }
    \label{f:fractions_w_literature}
\end{figure*}

To properly assess the performance of our classifier when comparing it with the literature results (which are taken as ground truth), we considered two factors. The first was to provide a comparison by selecting the best candidates as defined from the quality criteria in Sect. \ref{s:quality_cuts} (i.e., the final probability of > 0.66 and band completeness > 0.6). The second was to take into account the classification uncertainty by combining the full information from the probability distribution across all classes.

Since the classifier provides a probability for each of the classes, the probability of two (or more) classes may be similar. In that case, although a single final class is returned as the best prediction, it does not exclude others (e.g., probabilities of 0.45 and 0.40 for a RSG and YSG class, respectively). To account for this, we employed the categorical cross entropy (CCE) metric. It estimates the entropy contribution of each class based on the probability for the particular class returned from the classifier. This is multiplied with the one-hot encoder of the literature spectral type, which transforms a category to a vector with a length equal to the number of available classes. Only the class corresponding to the literature actually acts on the probability list. In other words, for each source:

\begin{align*}
CCE = - \sum_{k=1}^{N} [  y_\textrm{{true}} \times \ln( y_\textrm{{pred}} ) ],
\end{align*}

where $y_{\textrm{pred}}$ is the array of the probability returned by the classifier for each class, and $y_{\textrm{true}}$ is the vector that maps the literature type to the prediction list, both of them with dimensions of $1 \times N$ (in our case $N=7$ classes). For example, if our list of predictions has the following order: 

\begin{align*}
y_\textrm{{true}} = [ p_{\textrm{BSG}}, p_{\textrm{BeBR}}, p_{\textrm{GAL}}, p_{\textrm{LBV}},p_{\textrm{RSG}}, p_{\textrm{WR}}, p_{\textrm{YSG}} ].  
\end{align*}

If the literature provides a RSG classification, the one-hot encoded vector is

\begin{align*}
    y_{true}=[0,0,0,0,1,0,0].
\end{align*}

Then, the CCE considers only the probability for the RSG class, and reduces to 

\begin{align*}
CCE = - [ 1 \times \ln( p_{\textrm{RSG}})].
\end{align*}

In cases where the literature classification was broader (e.g., a  "blue" or "hot" star that is compatible with any of the predicted classes of BSG, LBV, WR, or BeBR), we used a soft one-hot encoding where the probability is equally split across all these classes, so that

\begin{align*}
CCE = - [ 0.25 \times \ln( p_{\textrm{BSG}} ) + 0.25 \times \ln( p_{\textrm{BeBR}}) \\ + 0.25 \times \ln( p_{\textrm{LBV}} ) + 0.25 \times \ln( p_{\textrm{WR}} ) ]. 
\end{align*}
 
By construction, lower CCE values indicate more secure classifications (the best being 0). However, this is far from true in practice. For this, we needed to define the threshold below which we considered our predictions secure. We already knew that our cutoff on probability was $p=0.66$. At the same time, such a probability actually meant that the prediction of the classifier was a single class with a relatively high probability (with probability at least double the probability of the next predicted class\footnote{In the case of two predicted classes, if the first (returned) class has a probability twice as large than the second one, we will have: $p_1+p_2=1 \Rightarrow 2 \times p_2 + p_1 = 1 \Rightarrow p_2=0.33$, so that $p1 = 0.66$.}). On the other hand, a random classifier with seven classes returns a probability of $1/7 \approx 0.14$. Therefore, we defined three regions that represented confidence in our predictions: a. "secure": $p > 0.66 \Rightarrow CCE < 0.42 $, b. "candidate": $ 0.14 < p \leq 0.66 \Rightarrow 0.42 \geq CCE > 1.95 $, and c. "uncertain": $ p \leq 0.14 \Rightarrow CCE \geq 1.95$. In that way, our estimate of robustness accounted for  the contribution from all classes, in conjunction with the information from the literature.  

We used the above approach in all sources for which a positional match between our catalog and the literature data was found within 1\arcsec. Sources without a match were excluded. Sources with ambiguous or uncertain literature classifications (such as the "composite" classification assigned to source B17 in WLM \citealt{Britavskiy2015}) were also excluded from further consideration, as they lack a reliable ground truth for validation. However, we retained these sources in the published catalog for completeness, but did not use them to assess classification performance.

We note here that when this comparison of spectral classification and the prediction from our photometric classifier is performed, variability may introduce an offset. Especially in cases of sources with significant variability (e.g., LBVs, RSGs), the epochs of spectroscopic and photometric observations (which are not concurrent in the optical and IR) may lead to a mismatch between the spectral type and the prediction. However, for larger populations (such as RSGs) we expect statistically consistent results.    

In Appendix \ref{s:appendix_literature_comparison} we provide in detail the statistics of matched sources per galaxy. In Fig. \ref{f:fractions_w_literature} we plot the success rate (corresponding to the number of secure and candidate classifications) over the subsample of literature sources, which matched the best candidates from our catalog with the distance and metallicity of each galaxy. We took extra care to deal with the errors in these fractions for two reasons. First, the fractions are bounded (0 to 1), so their error cannot extend beyond these values. Standard approaches of error estimate on the fractions (such as Wald or Wilson methods) yield symmetric errors, leading to values exceeding the bounds (e.g., close to 1). Second, the sample size is different in each galaxy. Based on the current sample size and the number of correct predictions (i.e., the total number of secure and candidate classifications, as defined previously), we get only one result (of the possible combinations) for the success rate, which follows a binomial distribution. To overcome all these and to avoid any Gaussian approximation, we followed a Bayesian approach (see Appendix \ref{s:appendix_errors} for a more technical overview). We estimated the posterior probability of the success rate based on the likelihood provided by the binomial distribution (given the sample size and the correct predictions in each case) and a prior. The prior corresponds to our knowledge of the distribution of probable success rates for our classifier. We considered two options for the prior, a uniform one (i.e., single success rate value) and a unimodal beta distribution (plots in Fig. \ref{f:fractions_w_literature} left and right correspondingly; only galaxies with literature data and good candidates, i.e., fulfilling the criteria of Sect. \ref{s:quality_cuts}, were plotted). This option translates to a success rate that has a single peak and drops off toward both bounds. We already had data for this from Paper I, in which we concluded that the accuracy was 0.83 for the M31 and M33 galaxies (on which the classifier was trained) and 0.70 for the test galaxies (i.e., IC 1613, WLM, and Sextans A). We therefore constructed a distribution with a mode at a mean value of 0.77 and a variance of 0.01\footnote{The difference between the two sample values is $\sim0.07\%$, which corresponds to a variance of 0.005. As this value is indicative of how the distribution drops from the peak value we opted to relax it, and double the value to 0.01, to allow for a smoother distribution to include more probable success rates.}. Given the likelihood and the prior, we constructed the posterior probability distribution per galaxy. From this we extracted basic statistical properties including asymmetric errors corresponding to the highest posterior density interval set at 95\%.

By examining the left panels of Fig. \ref{f:fractions_w_literature} we first notice that the success rate of the classifier drops with distance, although not significantly. This is expected, as the farther the galaxy is, the greater the level of confusion. The most extreme example is M83 for which we have, additionally, a small number of good candidates, but with relatively broader classifications from the literature (mainly \hii regions and emission stars). Interestingly, we also note that the classifier seems to be rather robust with metallicity. We achieve good results even at the lowest metallicity environments. There is an exception of three galaxies (namely NGC 7793, NGC 2403, and NGC 253) for which the success rate is lower than 0.4 (when using the uniform prior). These are the galaxies with the smallest number of classified samples for which we get very low success rates (e.g., one secure prediction out of the four sources in NGC 253, zero correct out of two in NGC 7793, and two secure and three candidates out of seven in NGC 2403). These cases are also sensitive to the low-number statistics, since a small change can lead to a significant change in the success rate. On the other hand, Sextans B has only two classified, of which we correctly matched and predicted only one. Hence, we ended up with a 100\% success rate, which was just a random realization of the binomial distribution. 

By incorporating the prior from the unimodal beta function that peaks at a mean success rate (as defined previously) we notice (see the right panel of Fig. \ref{f:fractions_w_literature}) that the values were updated and both extreme values (i.e., NGC 7793, NGC 253, NGC 2403, and Sextans B) are driven toward more probable (and realistic) values, within the 60--94\% success rate. The small decrease in the success rate with distance was still evident, but the performance of the classifier with respect to the metallicity flattened. M31 and M33 stand out in these plots because they both have the highest number of sources by far ($\sim430$ and $\sim380$, respectively, resulting in small errors) and the best success rates. The latter is expected since the classifier  has been trained on samples from these two galaxies.   

In summary, we conclude that our classifier performs exceptionally well even at lower metallicities (e.g., $0.07\,Z_\odot$ for Sextans A and Phoenix Dwarf), despite not being explicitly trained for them. There seems to be a greater dependence on distance (due to confusion), which makes it more efficient at distances smaller than 1.5 Mpc but with only a minor loss beyond 3 Mpc (where the furthest galaxies are located).

\subsection{Populations with metallicity}
\label{s:populations_with_metallicity}

\begin{figure*}[htbp]
    \centering
    \includegraphics[width=\textwidth]{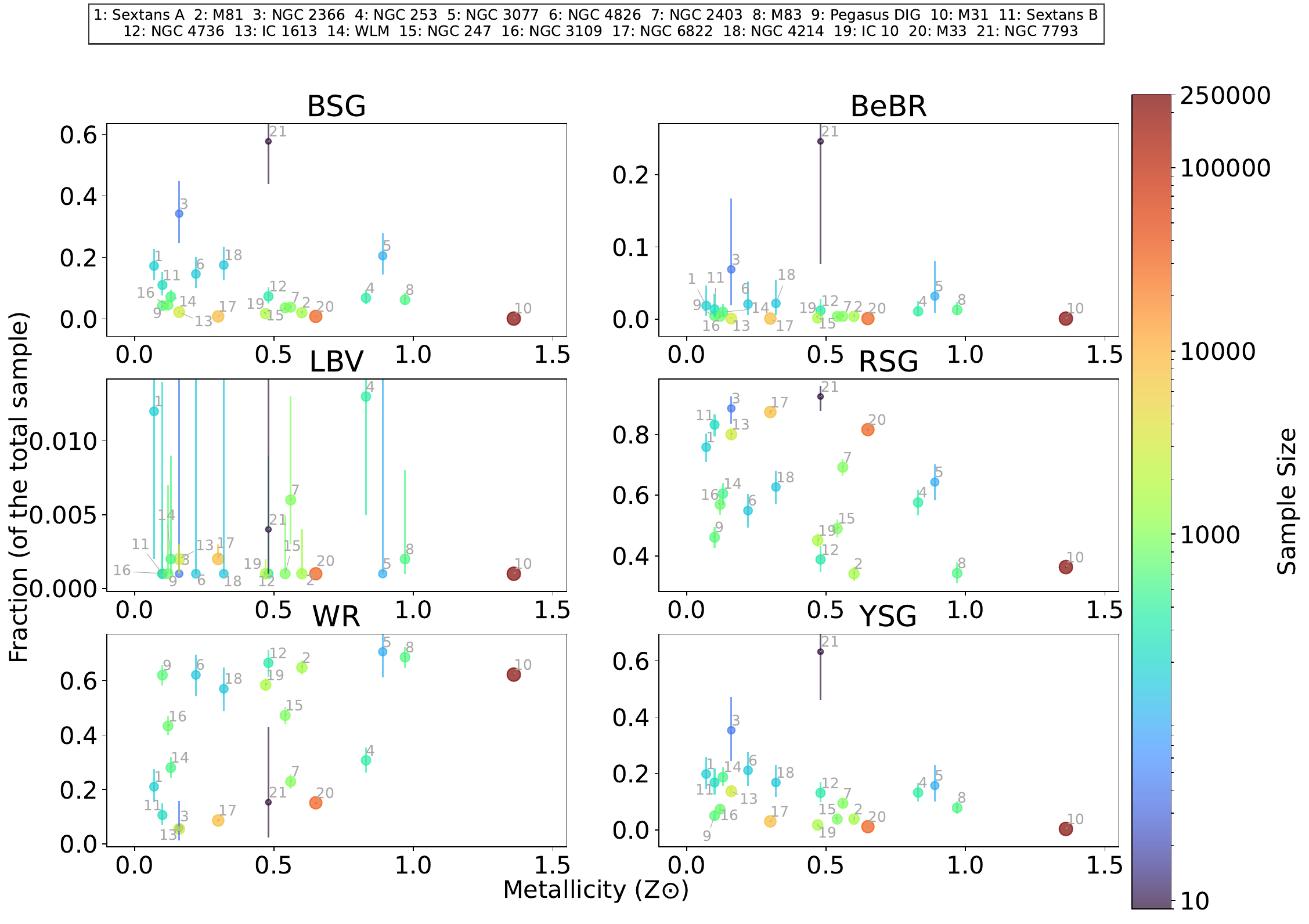}\\   
    \caption{Fraction of predicted population with metallicity per class. Despite the presence of large errors and important physical and observational biases, there are noticeable trends of each population with metallicity (see Sect. \ref{s:populations_with_metallicity} for details). The symbol size and color for each galaxy reflects the corresponding sample size. For clarity we assign each galaxy to an integer ID, shown in the legend at the top.}
    \label{f:fractions_w_Z}
\end{figure*}

In this section, we examine the trend of each class separately with metallicity. However, before presenting our results, we describe our biases in detail.

We repeat here that our source selection was based on \spitzer point source catalogs. That means that the catalog comprises of sources visible in the IR, which means that a certain fraction of some populations (e.g., OB main-sequence stars, stellar sources without dusty environments) are not included. Given this, we understand that the BSG class, for example, does not reflect the complete sample for any galaxy. This is evident if we compare the most recent catalog of OB stars in Sextans A by \cite{Lorenzo2022} with ours. We find matches for $\sim28\%$ (of 106 sources in total, within a 1\arcsec\,  search radius). Another factor influencing the fraction of populations that we can observe is the galaxy inclination. For example, NGC 55 is almost edge-on, which means that a (possibly significant) fraction is not visible.  In other cases, such as in Sextans A, which is a smaller and less crowded galaxy, we can pinpoint all sources. The difference between these two galaxies highlights that the galaxy's type and star formation history also have an important impact. A recent star-formation event will lead to the birth of new stars, which will probably (depending on the metallicity) lead to an increased number of WR stars. As time passes these populations (along with OB main sequence stars) will decrease, and in turn the populations of more evolved phases such as YSG and RSG will increase. 

Following the above physical reasons, there are several observational biases. One main constraint is the limitation in the pointings of \spitzer and its coverage, which affects the completeness. In \cite{Maravelias2023} we noted that in NGC 55 there are four known LBVs and our approach was successful in recovering two of them. The other two were not recovered due to their location, which lay outside the observed \spitzer fields. Therefore, an unknown fraction of all populations are affected. This is more critical in the rarest cases (such as for LBVs and BeBRs) while probably not significant for the most populous ones (such as RSGs and BSGs). Our classifier is based on the presence of photometry in certain \spitzer and Pan-STARRS bands. Additionally, many sources do not have photometry in all these bands, and consequently, the quality of their predictions did not pass our quality criteria. Although this compensates for not adding erroneous predictions, there is an important fraction of sources not considered in our analysis. 

Considering all the above caveats, Fig.~\ref{f:fractions_w_Z} presents the fraction of predictions per class over the total sample of only the best candidates. Our results are provided using the same approach for error determination described in the previous section (using the 95\% interval), but we use a separate prior for each class. The prior is derived from the unimodal beta function of each class, which peaks at the mean success rate and with a variance according to the results obtained in Paper I (cf., Table 5)\footnote{For the LBV class we used the fractions and corresponding variance derived from the support vector machine methods, the only nonzero result.}. We removed galaxies without any good matches (NGC 55, NGC 300, Phoenix Dwarf, NGC 1313, and NGC 5253).

Starting with LBVs, we cannot draw secure conclusions regarding their trend with metallicity given their small numbers and large errors. There is though indication that lower-Z environments may host more LBVs (e.g., Sextans A). 

The B[e] supergiants (BeBR) is another rare class of objects, for which there are not many confirmed cases in low metallicity galaxies (see e.g., in \citealt{Maravelias2023}). These sources are, by definition, very bright in IR due to the dust formation in their complex circumstellar rings, which means that we should be able to recover almost their total populations across all galaxies. Therefore, the slight increase with decreasing metallicity should be real. Whether or not we expect a higher number of such objects in low metallicity environments is currently unknown (because of the lack of their formation channels; see e.g., \citealt{Kraus2019a}). 

Metallicity has a significant impact on stellar winds of hot stars, which are mainly driven by iron. Therefore, decreasing metallicity leads to less efficient winds, resulting in fewer stars managing to expel their outer layers to become WR. This is depicted in the plot by the decrease in the WR fraction with metallicity, which is slightly more evident below 0.4 Z$_\odot$. Given that the fraction of false positives within WR is quite high, it is possible that their real fraction is even smaller than that depicted in the plot. Moreover, as they correspond to young massive stars, their populations also correlate with regions of recent star formation events (such as starburst NGC 4214; \citealt{Williams2011}). 

As the metallicity decreases, stars lose their angular momentum much slower, preserving in this way their initial rotational speeds longer and increasing the internal mixing (the stars become more compact and have higher temperature and density in their centers). Consequently, there seems to be an extension of the main-sequence phase (hydrogen burning at the core; \citealt{Georgy2013, Ekstrom2012}). Simultaneously, it is possible for He burning to occur much earlier (i.e., higher temperatures), before the RSG phase \citep{Yoon2008}. Those stars will spend more time as YSGs than RSGs. Moreover, the more massive and luminous RSGs have strong winds that lead to a fast stripping of their envelopes, resulting in post-RSGs that resemble BSGs or YSGs \citep{Massey2023, Zapartas2025}. Ultimately, these lead to more BSGs and YSGs present at any time than their corresponding numbers in higher metallicity galaxies. This is, indeed, observed in Fig. \ref{f:fractions_w_Z}.

The RSG population is not affected significantly (if at all) by metallicity (e.g., \citealt{Antoniadis2024, Antoniadis2025}). Therefore, we expect a rather flat trend or a slight drop in metallicity. However, this is not what we see in the RSG panel of Fig. \ref{f:fractions_w_Z}. We attribute this to observational bias, as we constructed our catalog based on \spitzer sources corresponding mainly to evolved and dusty sources. At lower metallicity galaxies, more stars spend their majority of the time in the BSG or YSG phase, which means that the number of these sources with dust will drop. Consequently, the ratios of RGSs over the total number of IR sources (which include fewer actually observed BSGs or YSGs as the metallicity decreases) lead to an upward trend.  

These intriguing findings reveal the impact of metallicity on massive star populations. They accentuate the need for more comprehensive studies to overcome current limitations imposed by the physical and observational biases of our approach.

\subsection{RSG luminosity functions}
\label{s:luminosity_functions_RSGs}

\begin{figure}[htbp]
    \centering
    \includegraphics[width=\linewidth]{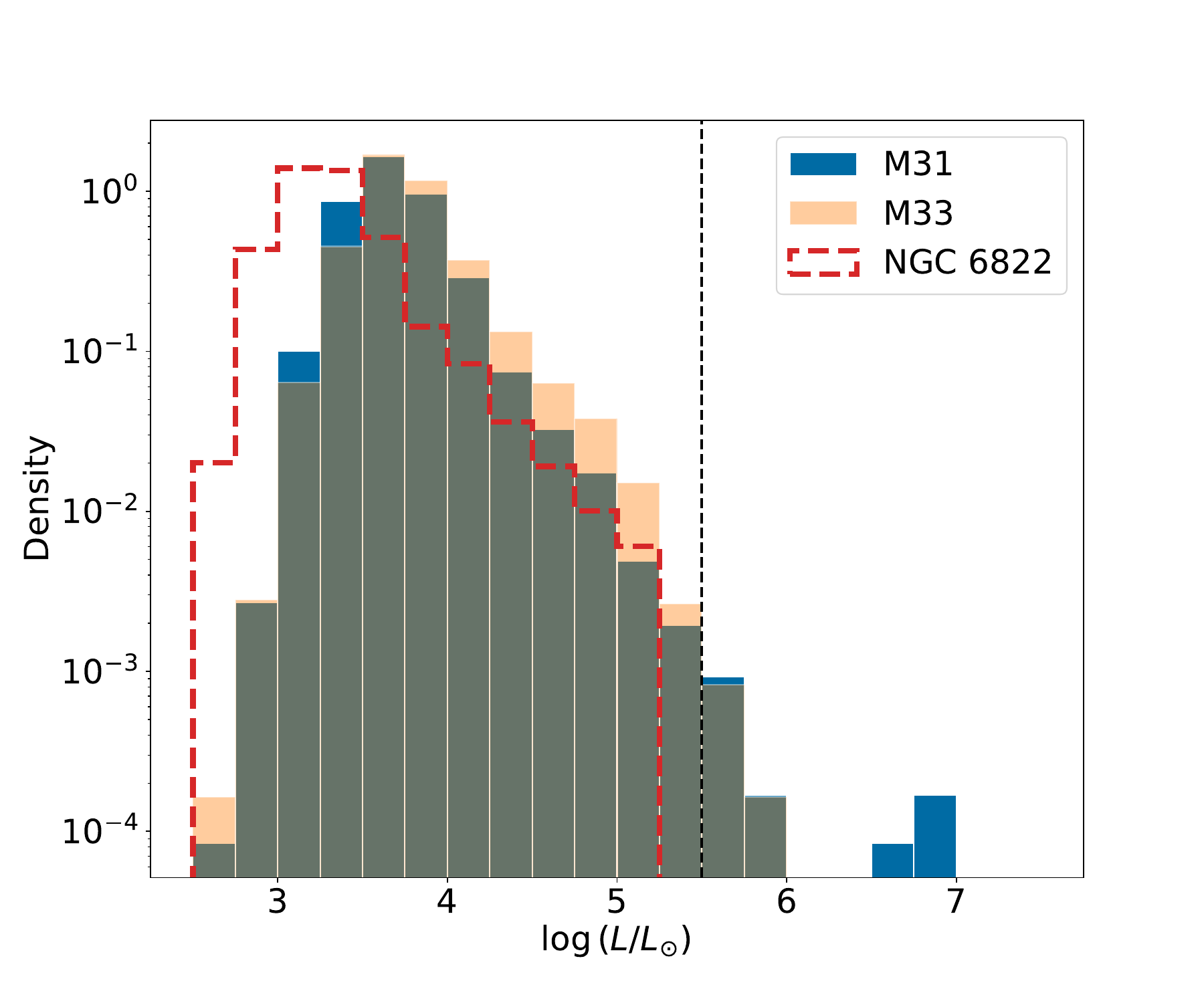}
    \caption{Luminosity functions for all sources identified as RSGs in M31, M33, and NGC 6822. We note the presence of some very luminous sources for M31 and M33 with \logl>5.5, indicated by a dashed line (see Sect.~\ref{s:luminosity_functions_RSGs}).}
    \label{f:RSG_luminosity_functions}
\end{figure}

\begin{table*}
  \centering
  \footnotesize
  \caption{Properties of luminous predicted RSGs in M31 and M33.} \label{t:luminous_rsgs}
  \begin{tabular}{lccccccccc}
  \hline
  \hline
  ID   &  \textit{Gaia}\_DR3\_id &  Parallax   &  pmRA  &  pmDec  &  Origi-l &  Fi-l    &  Fi-l       &  Band         & \logl  \\ 
       &                         &             &        &         &  Class    & Class     &  probability &  completeness &        \\
       &                         &  (mas)      &  (mas yr$^{-1}$) & (mas yr$^{-1}$) &  &  &              &               &  (dex)  \\
  \hline
M31-409650 & -- & -- & -- & -- & -- & RSG & 0.94 & 0.6 & 6.8 \\
M31-409552& -- & -- & -- & -- & -- & RSG & 0.67 & 1.0 & 6.8 \\
M31-409509& -- & -- & -- & -- & -- & RSG & 0.79 & 1.0 & 6.7 \\
M31-107& 381305998254302336 & $-$0.056 & 0.21 & 0.089 & M1I & RSG & 0.97 & 1.0 & 5.8 \\
M31-409640& 381275589889380352 & -- & -- & -- & -- & RSG & 0.87 & 1.0 & 5.8 \\
M31-171& -- & -- & -- & -- & -- & RSG & 0.97 & 1.0 & 5.7 \\
M31-155& 387318505783297792 & -- & -- & -- & -- & RSG & 0.98 & 1.0 & 5.7 \\
M31-184& 381285378106567680 & -- & -- & -- & -- & RSG & 0.96 & 1.0 & 5.7 \\
M31-409733& 381168662371881088 & $-$0.24 & 0.55 & $-$0.5 & -- & RSG & 0.87 & 1.0 & 5.7 \\
M31-144 & 381317508755717888 & -- & -- & -- & -- & RSG & 0.98 & 1.0 & 5.7 \\
M31-409898& 381172231487764352 & -- & -- & -- & -- & RSG & 0.98 & 1.0 & 5.7 \\
M31-183& 375276104678171392 & -- & -- & -- & -- & RSG & 0.98 & 1.0 & 5.7 \\
M31-212& 381285279321284352 & -- & -- & -- & -- & RSG & 0.9 & 1.0 & 5.6 \\
M31-409711& 381171067553249920 & $-$0.034 & 0.013 & 0.04 & -- & RSG & 0.91 & 1.0 & 5.6 \\
M31-201& 375296067688047616 & -- & -- & -- & -- & RSG & 0.70 & 1.0 & 5.5 \\
M31-409884& 381133306201068416 & 0.03 & $-$0.63 & $-$1.5 & -- & RSG & 0.76 & 1.0 & 5.5 \\
\hline
M33-37& 303288535788135680 & -- & -- & -- & -- & RSG & 0.84 & 1.0 & 5.8 \\
M33-197& 303267473268558208 & -- & -- & -- & -- & RSG & 0.98 & 1.0 & 5.6 \\
M33-34& 303365948277821952 & $-$0.18 & 0.099 & $-$0.18 & OB & RSG & 0.96 & 1.0 & 5.5 \\
M33-267& 315403264139899008 & -- & -- & -- & -- & RSG & 0.98 & 1.0 & 5.5 \\
M33-57& 303283553626082944 & $-$0.076 & 0.12 & 0.072 & RSG & RSG & 0.93 & 1.0 & 5.5 \\
M33-47& 303378072972636032 & $-$0.062 & $-$0.16 & $-$0.023 & RSG & RSG & 0.98 & 1.0 & 5.5 \\
  \hline
  \hline
  \end{tabular}

\end{table*}

Given that the highest success rate is obtained for RSGs, we opted to explore their luminosity functions for the galaxies with sufficient numbers, namely M31, M33, and NGC 6822. Humphreys-Davidson (HD) limit  \citep{Humphreys1979} is a well established empirical upper limit to the luminosity of massive stars. Hot stars are prohibited from moving to cooler temperatures, while LBVs can only sustain short excursions to cooler temperatures, which are considered unstable, and after intense or episodic mass loss they return to their original position in the HRD \citep{Higgins2020, Smith2026}. These stars do not pass through the RSG phase and they end up as WR stars. For lower mass stars (8-30 M$_\odot$) the evolution leads them all the way up the RSG phase. The HD limit marks the luminosity where substantial mass loss occurs (as they approach their Eddington limit) and is rarely found above that \citep[e.g.],[]{Davies2018, McDonald2022, Higgins2020}. Thus, RSGs either move to hotter positions in the HRD or die as core-collapse supernovae.  Although historically the HD limit \citep{Humphreys1979} is found at \logl$\sim5.8$ dex, more recent works lowered the limit to \logl$\sim5.5$ dex (\citealt{Davies2018} for the LMC and SMC; \citealt{McDonald2022} for M31). Hence, it is very intriguing to investigate predictions of RSGs above this limit.  

For this, we first need to estimate the luminosity for each source. We employed the bolometric correction for RSGs from \cite{Neugent2020} following their Sect. 5, given the $K_s-\textrm{magnitude}$ as provided by the UHS survey (using a 3" aperture for $JK_s$ photometry). To correct for extinction, we assumed a uniform value of $A_V=0.75$~mag as they did, but for all sources. When not applying additional extinction  for the brighter sources, our estimated luminosities are similar to those from \citep{Wang2021}, which are calculated by integrating the spectral energy distributions, a more robust method that takes into account the emission from the dust shell. For NGC~6822, we assumed $E(B-V)=0.25$~mag \citep{Massey2007}. The luminosities of RSGs in NGC~6822 are underestimated by 0.05 dex compared to the luminosities of the RSGs in common with \citet{Antoniadis2025}, which were calculated by integrating the spectral energy distributions. The results are presented in Fig.~\ref{f:RSG_luminosity_functions} (we note here that the uncertainty in this estimate due to the correction is 0.05 dex; \citealt{Neugent2020}).

From this plot, it is striking that there are several luminous sources above $5.5\,L_{\odot}$. In particular, we find 22 M31 and six M33 sources. In Table \ref{t:luminous_rsgs} we present them by including their \spitzer and \gaia DR3 IDs, their \gaia parallax and proper motions (whenever available), previous (original) and final (predicted) classifications, as well as their final probability, band completeness, and estimated luminosity. 

In M31, the three most luminous sources (close to \logl$\sim7$; IDs: M31-439614, M31-439351, and M31-439254) do not have a \gaia counterpart, which means that these could be foreground red stars. However, the next most luminous source (ID M31-350) is a confirmed M1 I RSG (J004428.48+415130.9), with a range of \logl$=5.43-5.64$ \cite{McDonald2022}. Such luminous sources may actually host a dense and complex circumstellar environment (similar to that observed in WOH G64; \citealt{Ohnaka2024, Munoz-Sanchez2025}) that can contribute up to 0.3 dex in luminosity. If we consider these uncertainties in our approach, then our estimate of \logl$=5.8$ could be corrected by 0.35 dex, making it consistent with values from the literature. Based on our \gaia foreground cleaning method, the parallax and proper motion measurements make it a strong M31 candidate. Therefore, it certainly provides an upper bound for the HD limit in M31. Below that there are an additional 12 sources (between \logl $\sim 5.8 - 5.5$), of which three have \gaia parallax and proper motion values making them the best candidates of high luminous RSGs\footnote{Among these sources is J003951.33+405303.7 from \cite{McDonald2022}, corresponding to M31-439741, which we predicted robustly as an RSG. Sources J004731.12+422749.1 (M31-1011) and J004539.99+415404.1 (M31-336) were also predicted as RSG, but we were unable to estimate their luminosity. For sources J004428.12+415502.9 and J004520.67+414717.3 we were unable to find a good match within a few arcseconds in our catalog, probably due to our different \gaia cleaning approach.}, another set of eight sources with \gaia photometry only (with at least one band measurement), and one source without any \gaia match (hence, more likely to be a foreground source). 

In M33, we find two sources (M33-179 and M33-520) at \logl$\sim6$). Since both of them lack \gaia parallax and proper motion measurements (but have photometric measurements), we cannot determine if they are genuine M33 members or not. These are followed by source M33-173 at \logl$\sim5.7$, for which full \gaia information exists. However, this is a known OB star \citep{Massey2016} and marks an erroneous prediction as an RSG. Another source (ID M33-646) has only \gaia photometry and, therefore, we can only tentatively consider it as a M33 RSG. The remaining two sources have full \gaia information and have been classified spectroscopically as RSGs. Therefore, they are the sources with the most robust data regarding their M31 membership status and spectral classification, and their derived luminosity sets a limit at \logl$\sim6$ (which could be lowered to \logl$\sim5.7$ if we consider the correction due to the circumstellar contribution).  

This exploration emphasizes the need for further detailed investigations into these luminous RSGs. This approach should be twofold. First, we should confirm their true nature as RSGs and accurately determine their luminosities through spectral energy distribution fitting. Second, we should be careful how to handle the stochastic effects due to the small size of the samples. Indeed, \citet{McDonald2022} comment that even though the observed $L_{\textrm{max}}$ is lower than the typical HD limit, when they examined the statistical significance of this $L_{\textrm{max}}$ they found that it is consistent within 3$\sigma$ with $L_{\textrm{max}}\sim5.8$, as derived from sampling a high number of sources from models. That means that our sources (with \logl$\geq5.8$) may be statistical outliers or foreground sources. Therefore, more careful work is needed in this direction. Larger samples and more precise luminosity measurements will definitely aid in exploring the HD limit. JWST is the perfect instrument to provide images that will decisively determine whether and how much these sources are contaminated.

\subsection{Dusty YSGs }

\begin{figure}[htbp]
    \centering
    \includegraphics[width=\linewidth]{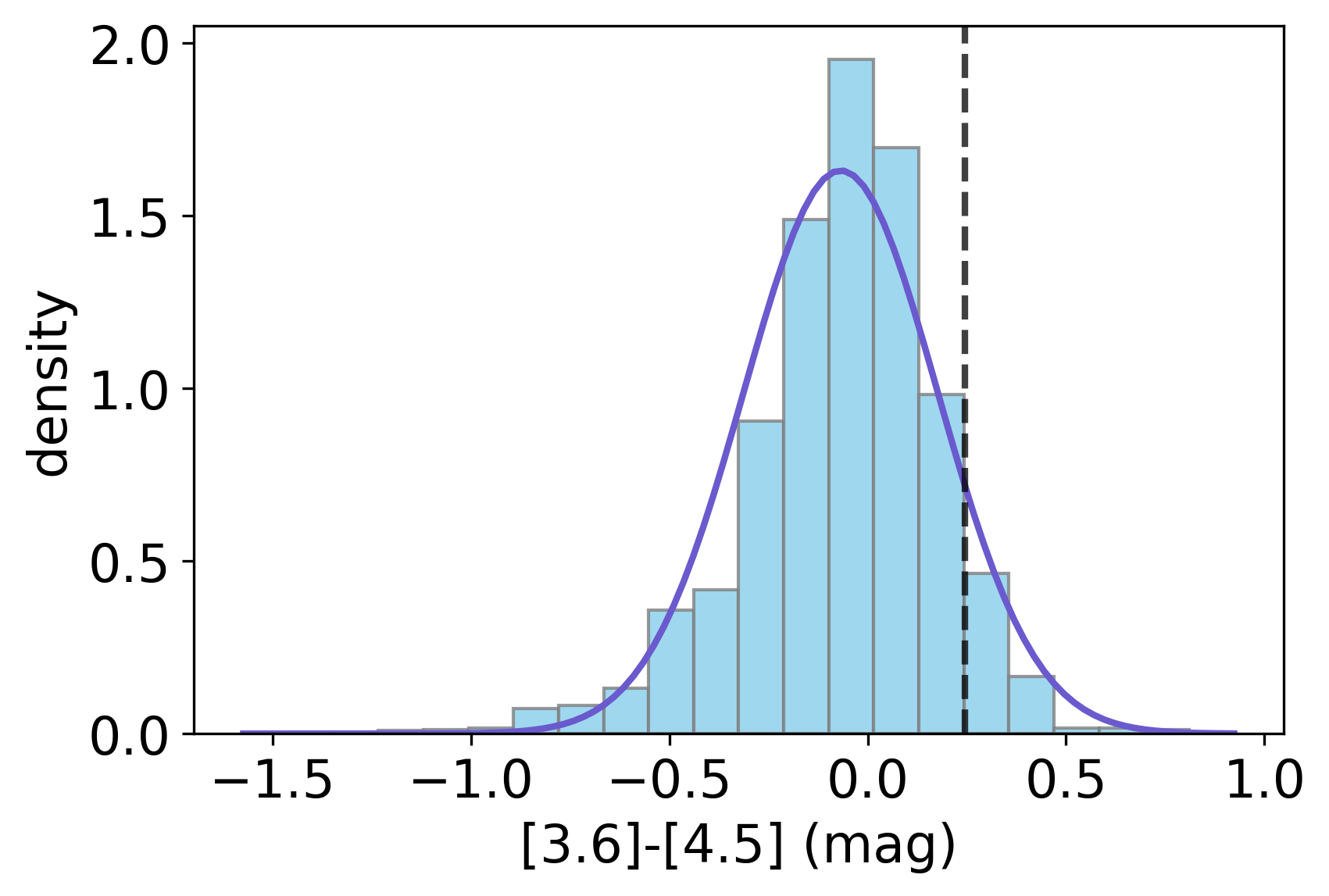}
    \caption{IR color ($[3.6]-[4.5]$) distribution of all (best candidate) sources predicted as YSGs. The violet curve corresponds to a Gaussian fit and the dashed black line indicates the $1\sigma$ threshold (0.24 mag) above which we identify sources as dusty YSGs.}
    \label{f:YSG_distribution}
\end{figure}

Since our catalog is based on IR photometry, we are particularly sensitive to dusty, evolved stars. This makes us especially effective at detecting and characterizing the population of dusty YSGs, which are likely the evolutionary descendants of luminous RSGs. The "RSG problem", i.e., the observed scarcity of luminous RSGs exploding as Type-II supernovae (e.g., \citealt{Smartt2009, Smartt2015, Davies2020, Rodriguez2022}) suggests that the most luminous RSGs may evolve to warmer temperatures, either by undergoing a blue loop phase (e.g., \citealt{Yoon2010, koumpia2020, Yang2023, Zapartas2025}) or due to binary interaction because RSG winds are not strong enough to strip RSGs (e.g., \citealt{Beasor2020, Decin2024, Antoniadis2024}). Due to their high mass-loss rates during the RSG phase, these YSGs are often surrounded by significant amounts of material where dust actively forms \citep{Gordon2019, Antoniadis2024, Decin2024}.

To investigate potentially interesting candidates in our sample, we considered only the YSGs fulfilling the quality criteria (as set for final probability and band completeness in Sect. \ref{s:quality_cuts}) in all galaxies. In Fig. \ref{f:YSG_distribution} we show the distribution of their IR color $[3.6]-[4.5]$. We fit this distribution with a Gaussian function and found a mean value of $-$0.07 mag (consistent with 0, which is the expected mean value) and a standard deviation of 0.24 mag. This value is also consistent with the value (0.25 mag) that has been used previously (e.g., \citealt{Kourniotis2017}) based on the LMC and SMC results (see \citealt{Bonanos2009, Bonanos2010}). Hence, we find 159 sources above the $1\sigma$ threshold (indicated in Fig. \ref{f:YSG_distribution}), which we identify as dusty YSGs, corresponding to a fraction of $\sim0.07$ with respect to their total number of 2071 sources. It is interesting to note that there are eight sources above the $2\sigma$ value (0.48 mag), namely M33-59527, M33-61917, M33-75218, M31-439225, M31-439242, M31-523223, WLM-19863, and NGC6822-30436), and another two (IC1613-14055 and WLM-234) above the $3\sigma$ threshold (0.72 mag). In Table \ref{t:dusty_ysgs} we present the IDs of all sources above the $1\sigma$ threshold. These are potentially yellow hypergiants that have suffered episodic mass loss and can provide a link to the "RSG problem." These sources need to be spectroscopically studied to verify their status.

\section{Summary and conclusions}
\label{s:summary}

In this study, we presented a comprehensive catalog of massive stars across 26 galaxies within 5 Mpc, leveraging a machine-learning classifier trained on optical and infrared photometry. Our classifier successfully classified 1,147,650 sources, of which 276,657 were deemed robust classifications based on probability and completeness criteria. Among these, we identified 120,479 RSGs, 2082 YSGs, 616 BSGs, 72 B[e] Supergiants, 150,151 WR stars, and 60 LBVs. A key result of our study is the effectiveness of the classifier across a broad metallicity range (0.07–1.36 Z$_\odot$), demonstrating its applicability even at low metallicities ($\sim0.1$ Z$_\odot$), despite not being explicitly trained for such environments. The classifier remains robust at distances $\le1.5$ Mpc, with only a slight decline beyond 3 Mpc due to the spatial resolution limits of \spitzer\!.

We investigated the effect of metallicity on different stellar populations, finding expected trends, such as a decrease in WR stars at lower metallicities and a relative increase in BSGs and YSGs. However, a number of selection biases, among which \spitzer\!'s sensitivity to dusty evolved stars, must be considered when interpreting these trends. We also identified 21 luminous RSGs (\logl$\ge5.5$), including six extreme RSGs in M31 (\logl$\ge6$), challenging the HD limit. Further investigation of these sources is necessary to confirm their nature and more accurately determine their luminosity, as well as to handle carefully the statistical perplexity due to the small sample sizes. In addition, 159 dusty YSGs were detected. These are optimal candidates of yellow hypergiants, key sources to understand the "RSG problem," and follow-up observations are vital. 

Our catalog provides the largest sample of machine-learning-classified massive stars in nearby galaxies, making it a crucial reference for future studies. It enables the identification of prime targets for spectroscopic follow-up, particularly luminous RSGs and YHGs, to further investigate their evolutionary pathways and the role of episodic mass loss in massive star evolution. Furthermore, we compiled a catalog of 5,273 spectroscopically confirmed sources for all 26 galaxies as derived from the literature. This offers a unique dataset for reference regarding spectral types for all massive stars and candidates known so far (including additional $\sim330$ other sources), for galaxies beyond the Milky Way and the Clouds. The accuracy of our classifier can be enhanced by incorporating additional spectral data and improving the handling of observational biases. Expanding this method with JWST photometry will allow the study of massive stars in more distant galaxies, providing deeper insights into their role in galactic evolution.

\hspace{0.5cm}

\section*{Data availability} Tables \ref{t:thecatalog}  and \ref{t:litcatalog} are only available in electronic form at CDS via anonymous ftp to cdsarc.u-strasbg.fr (130.79.128.5) or via http://cdsweb.u-strasbg.fr/cgi-bin/qcat?J/A+A/.\\ 

\begin{acknowledgements} The authors thank the referee for their useful comments. GM, AZB, KA, GMS, EC, SdW acknowledge funding support from the European Research Council (ERC) under the European Union’s Horizon 2020 research and innovation programme (``ASSESS'', Grant agreement No. 772086). EZ acknowledges support from the Hellenic Foundation for Research and Innovation (H.F.R.I.) under the “3rd Call for H.F.R.I. Research Projects to support Post-Doctoral Researchers” (Project No: 7933). 

\textit{Facilities:} This work has made use of data from the European Space Agency (ESA) mission {\it Gaia} (\url{https://www.cosmos.esa.int/gaia}), processed by the {\it Gaia} Data Processing and Analysis Consortium (DPAC, \url{https://www.cosmos.esa.int/web/gaia/dpac/consortium}). Funding for the DPAC has been provided by national institutions, in particular the institutions participating in the {\it Gaia} Multilateral Agreement. The Pan-STARRS1 Surveys (PS1) and the PS1 public science archive have been made possible through contributions by the Institute for Astronomy, the University of Hawaii, the Pan-STARRS Project Office, the Max-Planck Society and its participating institutes, the Max Planck Institute for Astronomy, Heidelberg and the Max Planck Institute for Extraterrestrial Physics, Garching, The Johns Hopkins University, Durham University, the University of Edinburgh, the Queen's University Belfast, the Harvard-Smithsonian Center for Astrophysics, the Las Cumbres Observatory Global Telescope Network Incorporated, the National Central University of Taiwan, the Space Telescope Science Institute, the National Aeronautics and Space Administration under Grant No. NNX08AR22G issued through the Planetary Science Division of the NASA Science Mission Directorate, the National Science Foundation Grant No. AST-1238877, the University of Maryland, Eotvos Lorand University (ELTE), the Los Alamos National Laboratory, and the Gordon and Betty Moore Foundation. The UHS is a partnership between the UK STFC, The University of Hawaii, The University of Arizona, Lockheed Martin and NASA.

\textit{Software:} This research made use of NumPy \citep{numpy2020}, pandas \citep{pandas2010, pandas2011}, matplotlib \citep{matplotlib}, imbalanced-learn \citep{imbalanced-learn}, scikit-learn \citep{sklearn}, SciPy \citep{scipy2020}, Astropy, a community-developed core Python package for Astronomy \citep{astropy2013, astropy2018}, Jupyter Notebooks \citep{jupyter}, TOPCAT, an interactive graphical viewer and editor for tabular data \citep{topcat}.

The NASA/IPAC Extragalactic Database (NED) is funded by the National Aeronautics and Space Administration and operated by the California Institute of Technology. This research has made use of NASA's Astrophysics Data System, and SIMBAD database, operated at CDS, Strasbourg, France. \end{acknowledgements}

%
%

\bibliographystyle{aa}
\bibliography{references}

\begin{appendix}
\clearpage
\onecolumn

\section{Detailed tables}

In this section, we present a collection of tables regarding our data processing and results.

\begin{table}[!h]
  \centering
  \caption{Number of sources per photometric survey and foreground selection criteria per galaxy (see Sect. \ref{s:gaia} for details.) 
   } \label{t:galaxy_statistics}
  \footnotesize
  \begin{tabular}{lrrrrrrrrrr}
  \hline
  \hline
  Galaxy   &  \textit{Spitzer}      &  PS-DR1  &  \textit{Gaia}-DR3           &  UHS  &  VHS  & Parallax/error & pmRA/error & pmDec/error  & Selected & Final \\ 

  \hline
WLM   & 14234 & 3331 & 399 & 0 & 3291 & $-$0.08+0.99 & 0.26$\pm$1.20 & $-$0.10$\pm$1.11 & 14091 & 13139 \\
NGC 55  & 8698 & 0 & 729 & 0 & 0 & $-$0.06+1.13 & 0.02$\pm$1.20 & 0.06$\pm$1.49 & 8524 & 8496 \\
IC 10  & 32901 & 4547 & 1931 & 6006 & 0 & $-$0.16+1.12* & 0.09$\pm$1.15* & $-$0.13$\pm$1.27* & 31673 & 29499 \\
M31  & 815811 & 410634 & 26332 & 387613 & 0 & $-$0.16+1.12 & 0.09$\pm$1.15 & $-$0.13$\pm$1.27 & 809142 & 809052 \\
NGC 247  & 13398 & 2470 & 621 & 0 & 0 & 0.01+0.85 & 0.21$\pm$0.98 & $-$0.03$\pm$1.14 & 13095 & 13095 \\
NGC 253 & 8734 & 1578 & 522 & 0 & 0 & 0.06+1.20 & $-$0.00$\pm$1.07 & 0.02$\pm$1.31 & 8409 & 8381 \\
NGC 300 & 20511 & 0 & 1400 & 0 & 11480 & 0.00+1.20 & 0.14$\pm$1.12 & $-$0.12$\pm$1.23 & 20161 & 20153 \\
IC 1613  & 28371 & 10364 & 1229 & 0 & 0 & $-$0.18+0.96 & 0.11$\pm$1.03 & $-$0.01$\pm$1.04 & 28245 & 26396 \\
M33 & 73206 & 52455 & 11049 & 47594 & 0 & $-$0.09+1.09 & 0.11$\pm$1.12 & 0.04$\pm$1.17 & 71847 & 71847 \\
Phoenix Dwarf  & 10831 & 0 & 499 & 0 & 1212 & 0.01+0.87 & 0.35$\pm$0.95 & $-$0.19$\pm$0.95 & 10703 & 10021 \\
NGC 1313 & 6156 & 6156 & 481 & 6156 & 6156 & 0.27+1.36* & 0.06$\pm$1.41* & 0.45$\pm$1.14* & 5970 & 5970 \\ 
NGC 2366 & 495 & 156 & 64 & 0 & 0 & $-$0.16+1.12* & 0.09$\pm$1.15* & $-$0.13$\pm$1.27* & 462 & 462 \\
NGC 2403 & 16644 & 3735 & 1517 & 0 & 0 & $-$0.01+1.43 & 0.26$\pm$1.64 & $-$0.05$\pm$1.47 & 15936 & 15910 \\
M81 & 28479 & 3894 & 1072 & 0 & 0 & $-$0.16+1.12* & 0.09$\pm$1.15* & $-$0.13$\pm$1.27* & 27895 & 27875 \\
Sextans B & 4914 & 1166 & 141 & 0 & 0 & $-$0.16+1.12* & 0.09$\pm$1.15* & $-$0.13$\pm$1.27* & 4852 & 4413 \\
NGC 3109 & 9474 & 2988 & 1069 & 0 & 0 & $-$0.09+0.93 & $-$0.01$\pm$1.06 & $-$0.05$\pm$1.18 & 8939 & 8935 \\
NGC 3077 & 2617 & 271 & 90 & 0 & 0 & $-$0.16+1.12* & 0.09$\pm$1.15* & $-$0.13$\pm$1.27* & 2548 & 2548 \\
Sextans A & 2888 & 880 & 219 & 0 & 355 & 0.03+0.84 & $-$0.20$\pm$0.98 & $-$0.12$\pm$0.94 & 2848 & 2693 \\
NGC 4214 & 1159 & 368 & 95 & 89 & 0 & $-$0.16+1.12* & 0.09$\pm$1.15* & $-$0.13$\pm$1.27* & 1149 & 1149 \\
NGC 4736 & 10043 & 1248 & 349 & 657 & 0 & $-$0.16+1.12* & 0.09$\pm$1.15* & $-$0.13$\pm$1.27* & 9861 & 9861 \\
NGC 4826 & 4659 & 480 & 149 & 306 & 0 & $-$0.16+1.12* & 0.09$\pm$1.15* & $-$0.13$\pm$1.27* & 4577 & 4575 \\
M 83 & 15020 & 2422 & 1396 & 0 & 3877 & $-$0.04+0.76 & 0.31$\pm$1.20 & 0.07$\pm$1.05 & 14132 & 14132 \\
NGC 5253 & 721 & 0 & 119 & 0 & 187 & $-$0.16+1.12* & 0.09$\pm$1.15* & $-$0.13$\pm$1.27* & 622 & 622 \\
NGC 6822 & 25599 & 18659 & 7061 & 0 & 15205 & 0.01+1.13 & $-$0.08$\pm$1.05 & $-$0.24$\pm$1.16 & 22483 & 22471 \\
Pegasus DIG & 11316 & 2234 & 251 & 0 & 0 & $-$0.16+1.12* & 0.09$\pm$1.15* & $-$0.13$\pm$1.27* & 11147 & 10530 \\
NGC 7793 & 5535 & 887 & 353 & 0 & 0 & 0.04+1.47 & 0.09$\pm$1.32 & -0.04$\pm$1.06 & 5433 & 5425 \\

  \hline
  \hline
  \end{tabular}

   \tablefoot{
    {\textit{Spitzer} data for galaxies IC 10, IC 1613, Pegasus DIG, Phoenix Dwarf, Sextans A and B, and WLM are derived from \citet{Boyer2015}; M33, M81, NGC 2403, NGC 247, NGC 300, NGC 6822, and NGC 7793 from \citet{Khan2015}; NGC 2366, NGC 253, NGC 4214, NGC 5253, and NGC 55 from \citet{Williams2016}; M31, M81, NGC 3077, NGC 4736, NGC 1313, and NGC 4826 from \citet{Khan2017}.  
    }
    \tablefoottext{*}{For those galaxies the parallax and proper motion criteria derived from M31 were used, as there were not enough data to determine these quantities by using their data only. 
    }

    }  

\end{table}

\begin{table}
  \centering
  \caption{Final source catalog with predicted classifications for all galaxies. 
  } \label{t:thecatalog}
  \footnotesize
  \begin{tabular}{lcccccccc}
  \hline
  \hline
  ID   &  RAJ2000  &  DEJ2000 &  \textit{Gaia}\_DR3\_ID  & ...  &  Final\_Prob & Final\_Class & Band\_Compl \\ 
      &  (deg)  &  (deg)   &    &  ...  &    &    &   & \\

  \hline
WLM-1 & 0.52017 & $-$15.44622 & -- & ... & 0.60202 & WR & 1.0 \\
WLM-2 & 0.52012 & $-$15.40681 & -- & ... & 0.64683 & WR & 0.4 \\
WLM-3 & 0.52012 & $-$15.36033 & -- & ... & 0.85906 & RSG & 1.0 \\
WLM-4 & 0.52012 & $-$15.46578 & -- & ... & 0.61820 & WR & 0.2 \\
WLM-5 & 0.52012 & $-$15.44331 & -- & ... & 0.64767 & WR & 0.2 \\
WLM-6 & 0.52012 & $-$15.41617 & -- & ... & 0.57403 & WR & 0.2 \\
WLM-7 & 0.52008 & $-$15.50714 & -- & ... & 0.61213 & WR & 0.2 \\
WLM-8 & 0.52008 & $-$15.46256 & -- & ... & 0.5356& WR & 0.2 \\
WLM-9 & 0.52008 & $-$15.56622 & -- & ... & 0.43782 & WR & 0.6 \\
WLM-10 & 0.52008 & $-$15.42581 & -- & ... & 0.61508 & WR & 0.2 \\
WLM-11 & 0.52008 & $-$15.43997 & -- & ... & 0.61310 & WR & 0.2 \\
WLM-12 & 0.52008 & $-$15.54617 & -- & ... & 0.55765 & WR & 0.2 \\
WLM-13 & 0.52008 & $-$15.56514 & -- & ... & 0.39619 & RSG & 0.2 \\
WLM-14 & 0.52004 & $-$15.53664 & -- & ... & 0.47621 & YSG & 0.6 \\
WLM-15 & 0.52004 & $-$15.55456 & -- & ... & 0.47340 & WR & 0.2 \\
WLM-16 & 0.52000 & $-$15.40131 & -- & ... & 0.42985 & WR & 0.6 \\
WLM-17 & 0.52000 & $-$15.46006 & -- & ... & 0.64583 & WR & 0.2 \\
WLM-18 & 0.52000 & $-$15.60042 & -- & ... & 0.61512 & WR & 0.2 \\
WLM-19 & 0.51996 & $-$15.45511 & -- & ... & 0.61603 & WR & 0.2 \\
WLM-20 & 0.51996 & $-$15.35928 & -- & ... & 0.48667 & RSG & 0.2 \\
  \hline
  \hline
  \end{tabular}

   \tablefoot{This table is available in its entirety in VizieR/CDS catalog tool. A portion is shown here for guidance regarding its form and content.}  

\end{table}

\begin{table}
  \centering
  \footnotesize
  \caption{IDs of sources predicted as YSGs and identified as dusty.} \label{t:dusty_ysgs}
  \begin{tabular}{llllll}
  \hline
  \hline
  ID   &  ID  &  ID   &  ID  & ID  & ID \\ 
  \hline
Sextans\_A-608 & NGC\_2403-10650 & M31-291254 & IC\_1613-2099 & IC\_1613-22555 & NGC\_6822-13118\\
M81-11303 & M83-2819 & M31-294149 & IC\_1613-2943 & IC\_1613-22602 & NGC\_6822-13324\\
NGC\_253-1293 & M83-4051 & M31-347036 & IC\_1613-3830 & IC\_1613-23280 & NGC\_6822-13679\\
NGC\_253-1478 & M83-7927 & M31-362279 & IC\_1613-4932 & IC\_1613-24144 & NGC\_6822-14299\\
NGC\_253-1870 & M83-13741 & M31-380446 & IC\_1613-5603 & IC\_1613-24766 & NGC\_6822-14322\\
NGC\_253-2676 & Pegasus\_DIG-6138 & M31-409496 & IC\_1613-5966 & IC\_1613-25537 & NGC\_6822-14942\\
NGC\_253-3150 & Pegasus\_DIG-8663 & M31-409503 & IC\_1613-6832 & WLM-69 & NGC\_6822-15128\\
M33-51717 & Pegasus\_DIG-8942 & M31-409505 & IC\_1613-6921 & WLM-179 & NGC\_6822-15582\\
M33-52312 & M31-3384 & M31-414106 & IC\_1613-7301 & WLM-188 & NGC\_6822-16756\\
M33-53306 & M31-3535 & M31-414142 & IC\_1613-8607 & WLM-9147 & NGC\_6822-17168\\
M33-55564 & M31-12298 & M31-451363 & IC\_1613-10328 & WLM-10681 & NGC\_6822-17404\\
M33-60369 & M31-45371 & M31-472507 & IC\_1613-10358 & WLM-13121 & NGC\_6822-17503\\
M33-62164 & M31-119542 & M31-479549 & IC\_1613-10781 & NGC\_3109-2169 & NGC\_6822-17521\\
M33-62367 & M31-129762 & M31-484559 & IC\_1613-11142 & NGC\_3109-2807 & NGC\_6822-18213\\
M33-63872 & M31-130114 & M31-581217 & IC\_1613-11194 & NGC\_3109-5242 & NGC\_6822-19404\\
M33-65643 & M31-132016 & M31-714077 & IC\_1613-11517 & NGC\_6822-1590 & NGC\_6822-19748\\
M33-65743 & M31-163495 & M31-749160 & IC\_1613-12270 & NGC\_6822-2820 & NGC\_6822-20276\\
M33-65915 & M31-173608 & M31-775787 & IC\_1613-14279 & NGC\_6822-3299 & NGC\_6822-20731\\
M33-68313 & M31-213432 & M31-791111 & IC\_1613-16613 & NGC\_6822-5305 & NGC\_6822-21191\\
M33-69050 & M31-217495 & Sextans\_B-1529 & IC\_1613-16662 & NGC\_6822-5378 & NGC\_6822-22201\\
M33-71044 & M31-227660 & Sextans\_B-1691 & IC\_1613-17410 & NGC\_6822-5581 & IC\_10-289\\
M33-71771 & M31-229975 & Sextans\_B-1745 & IC\_1613-19541 & NGC\_6822-7081 & IC\_10-858\\
NGC\_4826-41 & M31-246551 & Sextans\_B-3492 & IC\_1613-20037 & NGC\_6822-9860 & IC\_10-2758\\
NGC\_4826-109 & M31-256127 & NGC\_4736-847 & IC\_1613-20340 & NGC\_6822-10415 & IC\_10-19292\\
NGC\_4826-3560 & M31-261208 & NGC\_4736-4783 & IC\_1613-21452 & NGC\_6822-12447 &    \\
NGC\_2403-3659 & M31-272080 & IC\_1613-221 & IC\_1613-21676 & NGC\_6822-12948 &    \\
NGC\_2403-5096 & M31-273852 & IC\_1613-843 & IC\_1613-22503 & NGC\_6822-12966 &    \\
   \hline
  \hline
  \end{tabular}

\end{table}

\FloatBarrier

\section{Classified sources from the literature}
\label{s:appendix_ref_tables}

The following table contains references and the number of sources (per reference) obtained from the literature. We have taken care to remove any duplicates in the catalog, as well as to provide the most recent and precise classifications. We also excluded known foreground sources, but we kept any other source that was probably a member of the corresponding galaxy. \\

\begin{table}
    \centering
    \caption{References with their corresponding number of sources that contribute to our collected literature catalog.}
    \label{t:class_refs}
    \footnotesize
    \begin{tabular}{llr|llr}
  \hline
  \hline    
    Galaxy (Total) & Reference & Sources &
        Galaxy (Total) & Reference & Sources\\
    \hline
WLM (87) & \cite{Venn2003} & 2
& M33 (1547)$^{\ast\ast}$ & \cite{Martin2017} & 2 \\
         & \cite{Bresolin2006} & 38
&        & \cite{Humphreys2017} & 24 \\
         & \cite{Levesque2012} & 11
&        & \cite{Kourniotis2018} & 4 \\
         & \cite{Britavskiy2015} & 13
&        & \cite{Massey2019} & 10$^{6,8}$ \\
         & \cite{Britavskiy2019} & 1
&        & \cite{Neugent2019} & 46 \\
         & \cite{Maravelias2023} & 1
&        & \cite{Kraus2019a} & 7 \\
         & \cite{Bonanos2024} & 21
&        & \cite{Maryeva2020} & 2$^9$ \\
NGC 55 (279) & \cite{Castro2008} & 168
&            & \cite{Smith2020} & 1$^9$ \\
             & \cite{Patrick2017} & 11
&            & \cite{Neugent2021} & 82$^{10}$ \\
             & \cite{Maravelias2023} & 1
&            & \cite{Liu2022} & 18 \\
             & \cite{Bonanos2024} & 99
& Phoenix Dwarf (8) & \cite{Menzies2008} & 2 \\
IC 10 (137) & \cite{Massey2007} & 30
&            & \cite{Saviane2009} & 1 \\
            & \cite{Massey2007em} & 3
&            & \cite{Britavskiy2015} & 5 \\
            & \cite{Tehrani2017} & 29
& NGC 1313 (97) & \cite{Hadfield2007} & 80 \\
            & \cite{Britavskiy2019} & 6
&                & \cite{Bonanos2024} & 17 \\
            & \cite{deWit2025} & 69
& NGC 2366 (0) & n/a & 0 \\
M31 (1170)$^\ast$ & \cite{Massey2009} & 2
& NGC 2403 (68) & \cite{Humphreys2019} & 27 \\
                & \cite{Drout2009} & 18
&               & \cite{Bresolin2022} & 40 \\
                & \cite{Neugent2012} & 3
&               & \cite{Kaldybekova2023} & 1 \\
                & \cite{Massey2016} & 951
& M81 (82) & \cite{Kudritzki2012} & 26 \\
                & \cite{Gordon2016} & 81
&         & \cite{Khan2013} & 7 \\
                & \cite{Shara2016} & 1
&         & \cite{Humphreys2019} & 28 \\
                & \cite{Humphreys2017} & 6
&         & \cite{Gomez-Gonzalez2020} & 21 \\
                & \cite{Massey2019} & 17
& Sextans B (2) & \cite{Britavskiy2019} & 2 \\
                & \cite{Neugent2019} & 37
& NGC 3109 (127) & \cite{Evans2007} & 90 \\
                & \cite{Kraus2019a} & 11
&                & \cite{Flores-Duran2017} & 17$^{11}$ \\
                & \cite{Sholukhova2020} & 2
&                & \cite{Davidge2018} & 3 \\
                & \cite{Sarkisyan2022} & 1
&                & \cite{Maravelias2023} & 1 \\
                & \cite{Neugent2023} & 40
&                & \cite{Bonanos2024} & 16 \\
NGC 247 (64) & \cite{Solovyeva2020} & 2
& NGC 3077 (0) & n/a & 0 \\
              & \cite{Maravelias2023} & 1
& Sextans A (143) & \cite{Kaufer2004} & 5 \\
              & \cite{Bonanos2024} & 61
&                 & \cite{Britavskiy2014} & 2 \\
NGC 253 (81) & \cite{Comeron2003} & 1$^1$
&                 & \cite{Britavskiy2015} & 6 \\
              & \cite{Heida2015} & 1
&                 & \cite{Camacho2016} & 9 \\
              & \cite{Maravelias2023} & 1
&                 & \cite{Garcia2019} & 2 \\
              & \cite{Bonanos2024} & 78
&                 & \cite{Lorenzo2022} & 106 \\
NGC 300 (793) & \cite{Bresolin2002} & 66$^2$
&                 & \cite{Bonanos2024} & 13 \\
               & \cite{Schild2003} & 46
& NGC 4214 (0) & n/a & 0 \\
               & \cite{Crowther2007} & 10$^3$
& NGC 4736 (3) & \cite{Solovyeva2019} & 2 \\
               & \cite{Gazak2015} & 27
&               & \cite{Solovyeva2021} & 1$^{12}$ \\
               & \cite{Roth2018} & 504
& NGC 4826 (0) & n/a & 0 \\
               & \cite{Gonzalez-Tora2022} & 16$^4$
& M83 (241) & \cite{Hadfield2005} & 105$^{13}$ \\
               & \cite{Maravelias2023} & 2
&          & \cite{Bresolin2016} & 14 \\
               & \cite{Bonanos2024} & 122
&          & \cite{DellaBruna2022} & 66$^{13,14}$ \\
IC 1613 (86) & \cite{Kurtev2001} & 1
&          & \cite{Bonanos2024} & 56 \\
            & \cite{Bresolin2007} & 53
& NGC 5253 (0) & n/a & 0 \\
            & \cite{Herrero2012} & 1
& NGC 6822 (83) & \cite{Massey1998rsg} & 9 \\
            & \cite{Garcia2013} & 12
&               & \cite{Bianchi2001} & 1 \\
            & \cite{Britavskiy2019} & 3
&               & \cite{Massey2007em} & 2 \\
            & \cite{Chun2022} & 10$^5$
&               & \cite{Levesque2012} & 12 \\
            & \cite{deWit2025} & 6
&               & \cite{Patrick2015} & 10 \\
M33 (1547)$^{\ast\ast}$ & \cite{Massey1996} & 3
&               & \cite{deWit2025} & 49 \\
             & \cite{Massey1998rsg} & 52
& Pegasus DIG (9) & \cite{Massey2007em} & 1 \\
             & \cite{Massey1998wr} & 48$^6$
&             & \cite{Britavskiy2015} & 6 \\
             & \cite{Urbaneja2002} & 4
&             & \cite{Britavskiy2019} & 2 \\
             & \cite{Bruhweiler2003} & 2
& NGC 7793 (166) & \cite{Bibby2010} & 78$^{15}$ \\
             & \cite{Urbaneja2005} & 11$^7$
&              & \cite{Khan2013} & 3 \\
             & \cite{Massey2007em} & 12
&              & \cite{DellaBruna2020} & 5 \\
             & \cite{Neugent2011} & 2
&              & \cite{Wofford2020} & 36 \\
             & \cite{Drout2012} & 11
&              & \cite{DellaBruna2021} & 5 \\
             & \cite{Humphreys2014} & 4
&              & \cite{Maravelias2023} & 1 \\
             & \cite{Massey2016} & 1190
&              & \cite{Bonanos2024} & 38 \\
             & \cite{Gordon2016} & 12$^6$
& & & \\

     \hline
     \hline
    \end{tabular}
    \tablefoot{
    See notes in Appendix \ref{s:appendix_ref_tables}. 
}
 \end{table}

\noindent\textbf{Notes to Table~\ref{t:class_refs}.}
       
$^{\ast}$ Using the source list from Paper I. Updated classifications for three candidate to confirmed LBVs \citep{Sarkisyan2022,Sholukhova2020}, and 11 WR stars \citep{Neugent2023}. Removed two sources found as duplicates (original IDs: M31-957 and M31-982). Added WRs from \cite{Neugent2023}. 

$^{\ast\ast}$ Using IDs from Paper I.

$^1$ Spectral types are from \cite{Comeron2003}, while corrected coordinates from \cite{Comeron2003err}.

$^2$ Updated entries with the most recent $V$ magnitudes for 40 BSGs from \cite{Bresolin2005}. 

$^3$ Updated the list of sources presented in \cite{Schild2003}, by adding 10 more sources and removing 2 (as non-WR). 

$^4$ Updated spectral types for sources initially in \cite{Roth2018} (these entries were removed from the corresponding list). 

$^5$ Two sources (stars 4 and 12) were removed as they provided broader classifications rather than the classification provided by the older work of \cite{Britavskiy2019}. 

$^6$ Some $V$ magnitudes retrieved from \cite{Massey2007}.

$^7$ IDs OB 10-3 and UIT 136 refer to the same object. We kept the latest classification from that paper.

$^8$ Some $V$ magnitudes retrieved from \cite{Massey2016}.

$^9$ Confirmed LBV candidate from \cite{Massey2016}.

$^{10}$ Updated an uncertain YSG from \cite{Massey2016} to a RSG.
$^{11}$ $V$-band photometry from \cite{Pena2007}. \hii regions 41 -- 49 are knots or clumps in extended \hii regions, so they were excluded from the comparison. 

$^{12}$ Magnitude derived from \cite{Solovyeva2019}. 

$^{13}$ Assigned WR to all sources but kept the original classification within "[]". A magnitude value for the HeII $\lambda$4685 filter is provided wherever available; elsewhere a flag value of -99.0 is assigned. 

$^{14}$ Removed WR64 as duplicate to WR44, but WR35 and WR37 were kept as different (separation at 0.999\arcsec). 

$^{15}$ Removed all "not WR" objects (20), but kept "WN?" and "WC?" sources as WR candidates.  

\FloatBarrier

\section{Comparing the literature to predicted classes}
\label{s:appendix_literature_comparison}

In this appendix, we provide a detailed description of the comparison results between the literature and the classifier predictions per galaxy (except for NGC 2366, NGC 3077, NGC 4214, NGC 4826, and NGC 5253, for which no sources were found). For each case, we provide the number of sources found in the literature and the statistics in two distinct groups, one regarding the best (more reliable) results for sources that fulfilled the quality criteria as defined in Sect. \ref{s:quality_cuts}, and another for the whole set.

Regarding the comparison between the predicted class and the literature spectral type, we followed this matching algorithm: i. a predicted "RSG" was assumed to be compatible with RSG or candidates, sources with K or M types (even broader cases such as "early or late K or M"), carbon stars ("C-star","carbon"), broader labels such as "red" and "cool star"; ii. "WR" was compatible with any WR spectral type (WC, WN, etc.) including candidates; iii. "YSG" was compatible with any YSG, F or G (including "early or late F or G" labels), or "warm supergiant"; iv. "BSG" was compatible with any of the O, B, or A-type stars\footnote{The inclusion of A-type stars in this class follows the training of the method, see Paper I.} (and broader classes such as "early, mid, or late B"), OB or "hot supergiant"; v. "BeBR" was compatible with any source with a B[e] spectral type or candidate; vi. "LBV" was compatible with known and candidate LBVs; vii. "GAL" was compatible with any label indicative of a galaxy (e.g., "galaxy", "quasar", "AGN", and "QSO"), but we also considered them matching with carbon stars (pointing to elliptical galaxies where older stellar populations dominate); viii. broad spectral classifications from the literature such as "blue" or "hot", emission type stars (including labels as "Halpha star" and "OBem"), \hii regions and planetary nebulae ("PN") were considered compatible with any (predicted) class of "WR", "BeBR", "LBV", or "BSG"; ix. any other very coarse or doubtful case (such as "composite", "cluster", "star", "SNR", "neb", "symbiotic", "foreground") were considered uncertain. 

Regarding the matches between our catalog and the literature source, we considered a 1\arcsec search radius. These matches are referred to as "good matches". We point out here that the result of the process also depends on the accuracy of the coordinates taken from the literature.

From the total number of sources found in the literature for each galaxy, we first excluded those without matches within the search radius (as defined above). In addition, we also excluded ambiguous sources from the literature with very broad classifications (such as "Cluster", "Composite", "Star", "SNR", "Nebula"). From the remaining sources, we calculated the success rate by combining the number of secure and candidate sources over the total number of available sources. We provide detailed numbers for each of this group of sources.

In Table \ref{t:litcatalog} we provide the first few rows of the catalog that contains all sources with spectral classification from the literature. It includes all known massive stars and candidates, as well as another $\sim330$ sources of point sources (such as carbon stars, background galaxies, \hii regions, planetary nebulae, and clusters). We provide our own ID (with a preceding "lit-" to differentiate from our main catalog), an ID$_{lit}$ that corresponds to the ID provided in the corresponding work (if present). We provide the coordinates and the spectral type. For reference, a magnitude value in a specific filter is provided, as derived from the original paper, and in some cases corresponds to an indicative value (e.g., in variable sources such as in LBVs). When not available, a flagged value of -99.0 is given. Instead of the typical citation style, we give the full bibliography code as provided by NASA's Astrophysics Data System. 

\begin{table}[!h]
  \centering
  \caption{Compiled catalog of sources with spectral classifications from literature, including all massive stars and candidates, as well as additional point sources.
  } \label{t:litcatalog}
  \footnotesize
  \begin{tabular}{lllccccl}
  \hline
  \hline
  ID   &  LitID  &  RAJ2000  &  DEJ2000   &  SpecType  & Magnitude  & Filter  &  Bibcode \\
       &              &  (deg) &  (deg) &                 &  (mag)     &         &          \\
  \hline
lit-WLM-1 & [SC85b] 15 & 0.49812 & $-$15.49081 & A5 Ib & 18.1 & V & 2003AJ....126.1326V \\
lit-WLM-2 & [SC85b] 31 & 0.50262 & $-$15.47506 & A5 Ib & 18.4 & V & 2003AJ....126.1326V \\
lit-WLM-3 & A12 & 0.47183 & $-$15.47769 & B9 Ia & 17.98 & V & 2006ApJ...648.1007B \\
lit-WLM-4 & A11 & 0.49988 & $-$15.47200 & O9.7 Ia & 18.4 & V & 2006ApJ...648.1007B \\
lit-WLM-5 & A14 & 0.49825 & $-$15.49064 & A2 II & 18.43 & V & 2006ApJ...648.1007B \\
lit-WLM-6 & A9 & 0.48838 & $-$15.45503 & B1.5 Ia & 18.44 & V & 2006ApJ...648.1007B \\
lit-WLM-7 & A16 & 0.49125 & $-$15.50372 & A2 Ia & 18.44 & V & 2006ApJ...648.1007B \\
lit-WLM-8 & A19 & 0.50350 & $-$15.52111 & G2 I & 18.62 & V & 2006ApJ...648.1007B \\
lit-WLM-9 & B12 & 0.50267 & $-$15.47497 & A2 II & 18.77 & V & 2006ApJ...648.1007B \\
lit-WLM-10 & B13 & 0.47354 & $-$15.47497 & B1 Ia & 18.92 & V & 2006ApJ...648.1007B \\  
  \hline
  \hline
  \end{tabular}
   
    \tablefoot{This table is available in its entirety in VizieR/CDS catalog tool. A portion is shown here for guidance regarding its form and content.}

\end{table}

\subsection{WLM}

A total of 87 sources were collected from seven works in the literature, for which we obtained good matches (i.e., within 1\arcsec\,) for 62 sources ($\sim71\%$), resulting in 60 sources after removing two additional ambiguous literature classifications. We first checked the statistics for sources with the most reliable predictions, amounting to 30 sources (almost half of the sample). We obtained secure classifications for 86.7\% (26 sources) and uncertain classifications for 13.3\% (4 sources). Considering the whole sample (60 sources), we got 73.3\% (44 sources) secure classifications, 5.0\% (3) candidates, and 21.0\% (13) as uncertain. By combining the secure classifications and the candidates, we obtain a success rate of 78.3\%.

\subsection{NGC 55}

For NGC 55 we collected 279 sources from four different works. We removed 180 sources with no match at 1\arcsec\,  (equivalent to 64.5\% of the sample), and two other ambiguously classified objects. From the remaining 97 sources we did not have any reliable prediction according to the selection criteria we defined. Accounting for the whole sample,  we managed to obtain 1.0\% (1 source) as secure, 21.6\% (21) as candidates, and 77.3\% (73) as uncertain classifications. The total success rate is 22.7\%. We note here that almost all data in NGC 55 suffer from significant missing values (band completeness equal to 0.2), which means that the classifier cannot reach its top performance.

\subsection{IC 10}

The number of spectroscopically confirmed classified sources for IC 10 was 137 derived from five different papers. Of these sources, we did not find a good match for 79 sources (57.7\%), and we excluded another four sources with ambiguous classification. We had 11 sources passing the quality criteria, resulting in 63.6\% (7 sources) with secure classifications, 18.2\% (2 sources) as candidates, and 18.2\% (2 sources) as uncertain, with a total success rate of 81.8\%. By considering all (54) sources we got 37.0\% (20) secure sources, 24.1\% (13) candidates, and 38.0\% (21) uncertain ones. By combining the secure classifications and candidate numbers, we get a total success rate of 61.6\%.

\subsection{M31}

As the closest and largest galaxy the number of spectroscopically confirmed objects we managed to collect summed to 1170 sources (combining 13 different works). A bit more than half of them (57.4\%, 671 sources) do not have a good match. The majority of the remaining sources (382) passed the quality criteria, with 93.7\% (358) being secure predictions, about 1.0\% being candidates (4), and only 5.2\% being uncertain (20). Therefore, the total success rate was 94.8\% When considering all 499 sources, the fraction of secure predictions was 73.3\% (366), with additional 18.6\% (93) candidates and 8.0\% (40) uncertain cases. The total correct fraction is 92.0\%. These high fractions are not unexpected, since the classifier has been trained on (most) of these sources.

\subsection{NGC 247}

There are 64 sources in NGC 247 with secured spectra derived from three works. Of these, 17 sources ($\sim26.6$\%) do not have a good match within 1\arcsec, and another six sources have ambiguous classifications. There are 11 sources that fulfilled the selection criteria and we made secure predictions for nine of these (81.8\%), while the other two sources were uncertain (18.2\%).  The total success rate is 81.8\%. By considering all (41) sources we got secure classifications for 22.0\% (9 sources), 36.6\% (15) as candidates, and 41.5\% (17) as uncertain. By combining the secure and candidate classification, the total success rate  is 58.5\%.

\subsection{NGC 253}

We were able to gather 81 classified sources from four different works. Eighteen sources (22.2\%) were removed because of no match with any source from our catalog, and another 22 due to ambiguous classifications from the literature. Of the remaining 40 sources,  only 4 satisfied the quality criteria. From these we obtained secure predictions for 16.7\% (1 source), while 75.0\% (3) were uncertain. The success rate is 25.0\%. By considering the whole sample (40 sources), we  again got 2.5\% (1 source) as secure, 20.0\% (8) as candidates, and 77.5\% (31) as uncertain. The total success rate in this case is 22.5\%. Similarly to NGC~55, this galaxy also suffers from significant missing values (band completeness equal to 0.2), which results in the low success rate we obtained.

\subsection{NGC 300}

A total of 793 sources were collected from the literature. However, we had to remove the vast majority of them (553 sources, i.e., 69.7\%) because they did not have any good match within 1\arcsec\ of our catalog. We also excluded nine sources with ambiguous classification from the literature.  Due to the lack of Pan-STARRS coverage, there are no sources that can satisfy the band completeness criterion (>0.6). Considering all (231) sources, we got no secure classifications but only candidates (15.2\%, i.e., 35 sources). A fraction of 84.8\% (196) was predicted as uncertain. This leads to a success rate of 15.2\% solely from the candidates. Due to the many missing features in this galaxy, the fraction of uncertain predictions is quite high.

\subsection{IC 1613}

We collected 86 sources from seven different works. About 64\% of these (55 sources) do not have a good match with any of our catalog objects. Of the remaining 31 objects, 23 are good candidates (according to the selection criteria) and we managed to get 78.3\% (18 sources) secure predictions, 8.7\% candidates, and 13.0\% (3) as uncertain. The total success rate is 87.0\%. By considering all sources, we got 58.1\% (18) secure predictions, 19.4\% (6) candidates, and 22.6\% (7) uncertain  classifications. The total success rate is 77.4\%.

\subsection{M33}

For this galaxy, we have collected 1547 sources from 22 different works. About 61.3\% (948) did not have a good match with our catalog, and we excluded one ambiguous source. Of the remaining 598 sources, the majority (435) are of sufficient quality, and in particular 92.4\% (402) were secure predictions, along with 1.4\% (6) candidates, and 6.2\% (27) uncertain cases. This sums to 93.8\%. Considering all sources (598) we got 67.4\% (43 sources) secure predictions, 23.6\% (141) as candidates, and 9.0\% (54) as uncertain. The total fraction of secure and candidate sources is 91.0\%. Similarly to M31, we see high success rates since these sources lie in M33, which has been used for the training.

\subsection{Phoenix Dwarf}

Only eight objects with spectroscopic classifications were obtained from three different works in the literature. Unfortunately, the majority of them (75.0\%, i.e., six sources) do not have any match within the 1\arcsec\,  search radius. Moreover, these two remaining sources do not pass the quality criteria and they were predicted as uncertain. The success rate of 0\% is actually due to the small number statistics and the fact that the band completeness (in all of these cases) is very low (at 0.2), which means that most of the features are missing.

\subsection{NGC 1313}

We collected 97 sources from two different works. The majority of them (79.4\%, 77 sources) have no good match with our catalog, while another two have ambiguous classifications. Of the remaining 18 sources, there are no good candidates (i.e. that fulfill the selection criteria). We did not manage to securely predict any of them, but the majority (72.2\%, 13 sources) were predicted as candidates and 27.8\% (five sources) as uncertain. Therefore, the total success rate was 72.2\%.

\subsection{NGC 2403}

We identified 68 sources originating from two works \citep{Humphreys2019, Bresolin2022}, and an additional (tentative) case \citep{Kaldybekova2023}. Unfortunately, 53 ($\sim78\%$) of these sources (including the tentative case) have no good match with our catalog. Of the remaining 15 sources,  seven were of sufficiently good quality and we got 28.6\% (2) secure classifications, 42.9\% (3) as candidates, and 28.6\% (2) as uncertain. This results in a success rate of 71.4\%. When we considered all sources, we got 20.0\% (3) secure, 60.0\% (9) candidates, and 20.0\% (3) uncertain predictions. The total success rate was 80.0\%.  

\subsection{M81}

A total of 82 sources were collected from the four works in the literature (including a number of AGN/galaxies and \hii regions). No good match was found for the 67.1\% (55) sources. Of the remaining 27 sources, 12 passed the quality criteria, and we got 66.7\% (8) as secure classifications and 33.3\% (4) as uncertain, resulting in a success rate of 66.7\%. When considering all 27 sources we got 55.6\% (15) secure, 7.4\% (2) candidate\footnote{We note here that the catalog of \cite{Gomez-Gonzalez2020} includes two separate sources, namely WR-3 and WR-14, which are reported as different sources with a separation of 0.87". We opted not to remove any of them from our comparison list, although our search radius is 1\arcsec\, . Therefore, these two sources are matched with the same object from our catalog (with ID M81-471) and it is a duplicate result. The impact is not significant however, since this means a 3.6\% decrease in the candidates and the total success rate.}, and 37.0\% (10) uncertain classifications. The total success rate in this case was 63.0\%.  

\subsection{Sextans B}

Spectroscopic classifications hardly exist for this galaxy. There are a few objects studied by \citet{Britavskiy2019}, but the majority are foreground sources (and therefore excluded from further consideration). We were left with only two sources, of which one had a match within 1\arcsec\! . This source fulfilled the section criteria and was actually predicted correctly (leading to an obviously biased fraction of 100\%). 

\subsection{NGC 3109}

A total of 127 sources were gathered from five different works in the literature (including a number of \hii regions and planetary nebulae). No good match was found for the majority of the sources (74.9\%, 94 sources). We also excluded a single ambiguous classification. Of the remaining 32 sources, 14 passed the quality criteria, and we got 50.0\% (7) secure, 7.1\% (1) candidate, and 42.9\% (6) uncertain classifications. This resulted in a success rate of 57.1\%. When considering all available sources, we got 21.9\% (7) secure, 37.5\% (12) candidate, and 40.6\% (13) uncertain classifications. The total success rate in this case was 59.4\%. 

\subsection{Sextans A}

We collected 143 classified sources from seven different works. A significant fraction of 63.6\% of them (91 sources) were found more than 1\arcsec\,  away from our catalog sources. There were 14 sources with quality data, of which we got 57.1\% (8) secure, 21.4\% (3) candidate, and 21.4\% (3) uncertain classifications, accounting for a success rate of 78.6\%.  By considering all 52 sources, we got 15.4\% (8) secure, 42.3\% (22) candidate, and 42.3\% (22) uncertain classifications. The total success rate was 57.7\%.

\subsection{NGC 4736}

We only found three sources from the literature. Interestingly, all of them are candidate LBVs \citep{Solovyeva2019}, of which one has been confirmed based on photometric and spectroscopic variability \citep{Solovyeva2021}. Unfortunately, none of them had a good match with any of our catalog sources.

\subsection{M83}

We managed to obtain 241 sources from four different works. Unfortunately, the majority of them (185, 76.8\%) are not matched with our catalog sources. We also excluded another seven sources with ambiguous classification from the literature. Of the remaining 49 objects, only seven fulfilled the selection criteria. We got 28.6\% (2 sources) secure and 71.4\% (5) uncertain classifications (with a success rate of 28.6\%). Considering all (49) objects, we got 22.4\% (11) secure, 16.3\% (8) candidate, and 61.2\% (30) uncertain classifications. The total success rate was 38.8\%. This rate is quite low, because the majority of these sources correspond to \hii regions, which are consistent with a number of stellar sources (compatible with LBV, BSG, WR, and B[e]SG). When implementing categorical cross entropy all these classes are taken into account (with their corresponding probabilities), resulting at the end in less secure predictions.

\subsection{NGC 6822}

We obtained 83 sources from the literature (six different works). About 31\% (26 sources) did not have a good match, while another two have ambiguous classifications from the literature. Of the remaining 55 sources, 30 passed the quality criteria. We actually predicted almost all of them (96.7\%, 30 sources) securely, and only one source (3.2\%) as uncertain, i.e., a success rate of 96.7\%. This can be justified because the spectral types of these sources are RSG and we have the best success rate for this particular class. Considering all sources, we got 52.7\% (29 sources) as secure classifications, 23.6\% (13) as candidates, and 23.6\% (13) as uncertain. The total success rate was 76.4\%.

\subsection{Pegasus DIG}

We found nine sources from three different works. Two thirds of this sample (66.7\%, 6 sources) did not have any match within 1\arcsec\,. Of the remaining three objects, none fulfilled the selection criteria. Of these we got none with a secure classification, one candidate (33.3\%) and two uncertain (66.7\%) classifications. Therefore, the success rate was  33.3\%, but is based on a very small number of sources.

\subsection{NGC 7793}

We collected 166 classified sources from seven different works. A significant fraction (65.7\%, 109 sources) had no good match with our catalog, while an additional four sources have ambiguous classifications. From the remaining 53 sources, only two had quality data, and both were predicted as uncertain (leading to a 0\% success rate). Considering all sources, we got only 1.9\%, i.e., one source as secure classification, 30.2\% (16) as candidates, and 67.9\% (36)  as uncertain. The total success rate in this case was 32.1\%.

\section{A Bayesian approach to determine uncertainties in fractions}
\label{s:appendix_errors}

In this Appendix we describe the Bayesian approach used to determine the uncertainties in the fractions of the correct predictions. The likelihood is the probability of correctly classifying $k$ sources from the $n$ of those in a certain class, given the success probability $p$ of the classifier. Consequently, it is a binomial distribution:
\begin{equation}
    P(k|n,p) = \binom{n}{k} p^k (1-p)^{n-k}.
\end{equation}
The posterior probability of the success probability is given by Bayes' theorem:
\begin{equation}
    P(p|k,n) = \frac{P(k|n,p) P(p)}{P(k|n)},
\end{equation}
where $P(p)$ is the prior probability of the success rate, and $P(k|n)$ is the marginal likelihood, which can be treated as a normalization constant since it does not depend on $p$.

In this work, we report as point estimate the mode of the posterior, $m=\text{arg}\max\limits_p P(p|k,n)$. Allowing for custom priors, we perform the calculation numerically with a resolution of 0.001 in $p$, which is well below the resulting uncertainties by the range of values for $n$ and $k$ in our sample.

The reported uncertainty corresponds to the 68\% highest posterior density interval (HPDI), i.e., the smallest interval that contains 68\% of the posterior probability:
\begin{align}
    \text{HPDI} &= \left[ l, u \right], \\
    \int_{l}^{u} P(p|k,n) dp &= 0.68,
\end{align}
where the lower, $l$, and upper, $u$, bounds are such that they contain the mode, and have equal probability density except for cases where the interval includes the extrema of $p$, 0 or 1, in which it can only expand towards one tail. In our implementation, the HPDI is calculated by starting from the mode, and expanding the interval at lower or higher values of $p$, depending on the direction in which the posterior density is higher and the interval can expand if it reaches one of the extrema.

For the prior, we explore the uniform distribution, $P(p) = 1$, as well as the Beta distribution, which is the conjugate distribution of the binomial, and is often employed as prior on probabilities. For the latter case, since we aim to use a prior with location and spread representing previous results, we constructed an algorithm that finds the parameters $\alpha$ and $\beta$ of the Beta distribution that result in the desired mode and variance. We do this by numerically solving for the parameters in the system of equations:
\begin{align}
    \text{mode} &= \frac{\alpha-1}{\alpha+\beta-2}, \\
    \text{variance} &= \frac{\alpha\beta}{(\alpha+\beta)^2(\alpha+\beta+1)},
\end{align}
where the variance is less than $\frac{1}{12}$ to ensure unimodality.

The code for the Bayesian analysis and the search for the unimodal Beta distribution, as well as the documentation, are provided in the following open source repository: \url{https://github.com/kkovlakas/gaussfree}.

\end{appendix}

\end{document}